\pdfoutput=1
\documentclass[bibyear]{aa}

\usepackage{graphicx}
\usepackage{color}
\usepackage{epstopdf}
\usepackage{amssymb}
\usepackage[varg]{txfonts}
\begin{document} 

\title{High-resolution imaging of the molecular outflows in two mergers: \object{IRAS~17208-0014} and \object{NGC~1614} \thanks{Based on observations carried out with the IRAM Plateau de Bure Interferometer. IRAM is supported by INSU/CNRS (France), MPG (Germany), and IGN (Spain).}}
            
             \author{S.~Garc\'{\i}a-Burillo \inst{1}
                        \and
           F.~Combes\inst{2}            
                        \and
           A.~Usero\inst{1}                
                        \and
           S.~Aalto\inst{3}
                \and                    
           L.~Colina\inst{4,5}   
                 \and                   
           A.~Alonso-Herrero\inst{6}            
                        \and
          L.~K.~Hunt\inst{7}
                         \and
          S.~Arribas\inst{4,5}
                        \and    
          F.~Costagliola\inst{8}         
                        \and
          A.~Labiano\inst{9}
                \and
          R.~Neri\inst{10}
                \and
        M.~Pereira-Santaella\inst{4}
                \and
          L.~J.~Tacconi\inst{11}
                  \and
          P.~P.~van der Werf\inst{12}     
          }
   \institute{
          Observatorio Astron\'omico Nacional (OAN-IGN)-Observatorio de Madrid, Alfonso XII, 3, 28014-Madrid, Spain 
                          \email{s.gburillo@oan.es}               
      \and
      Observatoire de Paris, LERMA, CNRS, 61 Av. de l'Observatoire, 75014-Paris, France 
         \and    
         Department of Earth and Space Sciences, Chalmers University of Technology, Onsala Observatory, 439 94-Onsala, Sweden        
         \and
        Centro de Astrobiolog\'{\i}a (CSIC-INTA), Ctra de Torrej\'on a Ajalvir, km 4, 28850 Torrej\'on de Ardoz, Madrid, Spain 
        \and
         ASTRO-UAM, Universidad Aut\'onoma de Madrid (UAM), Unidad Asociada CSIC, Madrid, Spain         
         \and
          Instituto de F\'{\i}sica de Cantabria, CSIC-UC, E-39005 Santander, Spain.   
         \and
         INAF-Osservatorio Astrofisico di Arcetri, Largo Enrico Fermi 5, 50125-Firenze, Italy     
         \and
          Instituto de Astrof\'{\i}sica de Andaluc\'{\i}a (CSIC), Apdo 3004, 18080-Granada, Spain 
        \and     
        Institute for Astronomy, Department of Physics, ETH Zurich, CH-8093 Zurich, Switzerland     
         \and    
         Institut de Radio Astronomie Millim\'etrique (IRAM), 300 rue de la Piscine, Domaine Universitaire de Grenoble, 38406-St.Martin d'H\`eres, France 
         \and
         Max-Planck-Institut f\"ur extraterrestrische Physik, Postfach 1312, 85741-Garching, Germany 
        \and
        Leiden Observatory, Leiden University, PO Box 9513, 2300 RA Leiden, Netherlands 
        }
         
        \date{Received ---; accepted ----}
             
 
  \abstract
  {Galaxy evolution scenarios predict that the feedback of star formation and nuclear activity (AGN) can drive the transformation of gas-rich spiral mergers into (ultra) luminous infrared galaxies and, eventually, lead to the build-up of QSO/elliptical hosts.}
   {We study the role that star formation and AGN feedback have in launching and maintaining the molecular outflows in two starburst-dominated advanced mergers, \object{NGC~1614} ($D_{\rm L}=66$~Mpc) and \object{IRAS~17208-0014} ($D_{\rm L}=181$~Mpc), by analyzing the distribution and kinematics of their molecular gas reservoirs. Both galaxies present evidence of outflows in other phases of their ISM.}
   {We used the Plateau de Bure interferometer (PdBI) to image the CO(1--0) and CO(2--1) line emissions in \object{NGC~1614} and \object{IRAS~17208-0014}, respectively, with high
spatial resolution (0$\farcs$5--1$\farcs$2). The velocity fields of the gas were analyzed and modeled to find the evidence of molecular outflows in these sources and characterize the mass, momentum, and energy of these components.}
   {While most ($\geq95\%$) of the CO emission stems from spatially resolved ($\sim2-3$~kpc-diameter) rotating disks, we also detect in both mergers the emission from high-velocity line wings that extend up to $\pm$500--700~km~s$^{-1}$, well beyond the estimated virial range associated with rotation and turbulence. The kinematic major axis of the {\it \emph{line-wing}} emission is tilted by $\sim90^{\circ}$ in \object{NGC~1614} and by $\sim180^{\circ}$ in \object{IRAS~17208-0014} relative to the major
axes of their respective rotating disks. These results can be explained by the existence of non-coplanar molecular outflows in both systems: 
the outflow axis is nearly perpendicular to the rotating disk in \object{NGC~1614}, but it is tilted relative to the angular momentum axis of the rotating disk in \object{IRAS~17208-0014}.}
   {In stark contrast to \object{NGC~1614}, where star formation alone can drive its molecular outflow, the mass, energy, and momentum budget requirements of the molecular outflow in \object{IRAS~17208-0014} can be best accounted for by the existence of a so far undetected (hidden) AGN of $L_{\rm AGN}\sim7\times10^{11}~$L$_{\sun}$. The geometry of the molecular outflow in \object{IRAS~17208-0014} suggests that the outflow is launched by a non-coplanar disk that may be associated with a buried AGN in the western nucleus.}

   \maketitle   
%

\section{Introduction}

 Different theoretical models and numerical simulations have long foreseen that massive gas outflows powered by active galactic nuclei (AGN) and/or nuclear starbursts could link the growth of black holes and
their hosts during the history of galaxy evolution, as inspired by the ground-breaking work of Silk \& Rees~(\cite{Sil98}). Outflows could prevent galaxies from becoming overly massive and help regulate the fueling of both the star formation and the nuclear activity (e.g., King~\cite{Kin03}; Di Matteo et al.~\cite{DiM05, DiM08}; 
Ciotti \& Ostriker~\cite{Cio07};  Narayanan et al.~\cite{Nar08}; Silk \& Nusser~\cite{Sil10};  Ishibashi  \& Fabian~\cite{Ish12}; Faucher-Gigu{\`e}re \& Quataert~\cite{Fau12}; Hopkins et al.~\cite{Hop12, Hop14};  Zubovas \& King~\cite{Zub12, Zub14}; Gabor \& Bournaud~\cite{Gab14}).

There is mounting observational evidence of gas outflows  in 
different populations of starbursts and active galaxies, including ultra luminous infrared galaxies (ULIRGs), radio galaxies, quasars, and Seyferts.
The outflow phenomenon concerns virtually all the phases of the interstellar medium (ISM), including the ionized gas (e.g., Heckman et al.~\cite{Hec90}; Colina et al.~\cite{Col91}; Martin~\cite{Mar98}; Holt et al.~\cite{Hol08}; Crenshaw et al.~\cite{Cre10}; Rupke \& Veilleux~\cite{Rup13}; Arribas et al.~\cite{Arr14}), the neutral atomic medium  (e.g., Heckman et al.~\cite{Hec00}; Rupke et al.~\cite{Rup02, Rup05a, Rup05b, Rup05c}; Morganti et al.~\cite{Mor05}; Martin~\cite{Mar05, Mar06}; Chen et al.~\cite{Che10}), as well as the molecular gas phase (e.g., Feruglio et al.~\cite{Fer10}; Fischer et al.~\cite{Fis10}; Sturm  et al.~\cite{Stu11}; Alatalo et al.~\cite{Ala11, Ala15}; Chung et al.~\cite{Chu11}; Aalto et al.~\cite{Aal12}; Dasyra \& Combes~\cite{Das12}; Maiolino et al.~\cite{Mai12}; Combes et al.~\cite{Com13}; Morganti et al.~\cite{Mor13}; Veilleux et al.~\cite{Vei13}; Cicone et al.~\cite{Cic12, Cic14, Cic15}; Garc\'{\i}a-Burillo et al.~\cite{Gar14}; Gonz{\'a}lez-Alfonso et  al.~\cite{Gon14}; Davies et al.~\cite{Dav14}; Emonts et al~\cite{Emo14}; Sakamoto et al.~\cite{Sak14}). This multiwavelength evidence underlines  that the overall impact of outflows can only be quantified if all of their ISM phases are studied in different populations of galaxies.

 Although ionized gas flows are seen to be typically a few 100~km~s$^{-1}$ faster than their atomic or molecular counterparts,  the neutral medium is seen to evacuate mass ($M_{\rm out}$) and energy ($E_{\rm out}$) at comparably higher rates ($M_{\rm out}(neutral)\sim(10-100) \times M_{\rm out}(ionized)$ and
 $E_{\rm out}(neutral)\sim(1-100) \times E_{\rm out}(ionized)$; e.g., Rupke \& Veilleux~\cite{Rup13}). Since the molecular component is the most massive phase of the neutral
 ISM in the central kiloparsec regions, the study of molecular outflows is therefore crucial to gauging their global impact on galaxy evolution.

The high spatial resolution and sensitivity capabilities, allied with the enhanced bandwidth of current mm-interferometers have revealed the existence of cold molecular outflows at kiloparsec scales in a growing number of (U)LIRGs. Outflows are detected by the low-level broad
line wings (up to $\sim \pm~1000$~km~s$^{-1}$) identified mainly, yet not exclusively, in CO lines (Feruglio et al.~\cite{Fer10, Fer13}; Cicone et al.~\cite{Cic12, Cic14}). Tracers more specific to the dense molecular gas, such as low and mid-J rotational lines of HCN and HCO$^+$, have been used to study the outflow in the prototypical AGN-dominated ULIRG Mrk~231 (Aalto et al.~\cite{Aal12, Aal15a}). 

Cicone et al.~(\cite{Cic14})  present new CO observations of seven ULIRGs and report detecting molecular outflows in four out of seven of their targets. The outflows, extending on kiloparsec scales, present mass load rates (d$M$/d$t$) of several 100~$M_{\sun}$~yr$^{-1}$. Based on their compiled sample of 19 sources, they found that star formation can drive outflows with d$M$/d$t$ values up to two to four times the value of the star formation rate ($SFR$). However, the presence of an AGN, even if weak ($L_{\rm AGN}/L_{\rm bol} < $~0.10),  can strongly boost an outflow with   d$M$/d$t \propto L_{\rm AGN}$. In the most extreme quasars ($L_{\rm AGN}/L_{\rm bol} \geq 0.80$), outflow rates can be up to $100 \times SFR$, and therefore have a decisive  impact on the fueling of the star formation and nuclear activity of their hosts.

In an attempt to study the specific role that star formation versus AGN feedback have in launching and maintaining molecular outflows in mergers, we used the Plateau de Bure interferometer (PdBI) to image the CO emission in two starburst-dominated (U)LIRGs, \object{NGC~1614} and \object{IRAS~17208-0014}, with high spatial resolution
(0$\farcs$5--1$\farcs$2). The two targets present previous evidence of an outflow in several gas tracers, including the ionized, the atomic, and the molecular ISM (\object{NGC~1614}: Bellocchi et al.~\cite{Bel12}; \object{IRAS~17208-0014}: Sturm et al.~\cite{Stu11}; Rupke \& Veilleux~\cite{Rup13}; Arribas et al.~\cite{Arr14}). Both systems are classified as mergers in an advanced stage of interaction, and they host massive episodes of star formation. Interestingly,  although recently questioned by Aalto et al.~(\cite{Aal15b}) in the case of \object{IRAS~17208-0014}, there has been no evidence to date for an energetically  relevant AGN in these mergers at their late stage of evolution ($L_{\rm AGN}/L_{\rm bol} < 0.1-11\%$; e.g., Gonz\'alez-Mart\'{\i}n et al.~\cite{Gon09}; Rupke \& Veilleux et al.~\cite{Rup13}; Herrero-Illana et al.~\cite{Her14}).  The absence of any clear signs of ongoing nuclear activity makes the choice of these galaxies optimum in order to test the canonical evolutionary scenario where gas-rich spiral mergers lead first to a (U)LIRG and, eventually, under the action of AGN feedback to the build-up of QSO/elliptical hosts (e.g., Sanders et al.~\cite{San88}; Veilleux et al.~\cite{Vei09}). Furthermore, the high resolution of the PdBI observations used in this work are instrumental in constraining the geometry, as well as the mass, momentum, and energy budgets of the molecular outflows unveiled in \object{NGC~1614} and \object{IRAS~17208-0014}.

We describe in Sect.~\ref{Targets} the two targets that are the subject of this study. Section~\ref{Obs} describes the PdBI observations.  Section~\ref{Results} presents the CO line maps obtained and discusses the observational evidence of molecular outflows in both sources. Section~\ref{out-properties} describes the basic properties of the molecular outflows. We discuss these results in the context of the molecular outflow phenomenon in (U)LIRGs in Sect.~\ref{out-comparison}. The main conclusions of this work are summarized in Sect.~\ref{Summary}.

\section{Targets}\label{Targets}

\subsection{\object{NGC~1614}}

\object{NGC~1614} ($D_{\rm A}=64$~Mpc, $D_{\rm L}=66$~Mpc; $1\arcsec=310$~pc) is an H{\small II}-classified LIRG ($L_{\rm IR}=4.5 \times 10^{11}~L_{\sun}$) in an advanced stage 
of interaction. The bolometric luminosity of \object{NGC~1614} is mainly powered by an extreme star formation episode with an integrated $SFR$ of 
$\sim 50 M_{\sun}$yr$^{-1}$ (Alonso-Herrero et al.~\cite{Alo01}, U et al.~\cite{U12}). Star formation is fed by a $\sim3 \times 10^{9}~M_{\sun}$ 
molecular gas reservoir (K\"onig et al.~\cite{Kon13}). Numerical simulations suggest that the burst was triggered after a minor merger 
(V\"ais\"ainen et al.~\cite{Vai12}). In their multiwavelength study, Herrero-Illana et al.~(\cite{Her14}) concluded that there is no evidence of an AGN in the nucleus of \object{NGC~1614} down to an AGN-to-bolometric luminosity ratio of $<10\%$. In agreement with this picture,  not detecting the nuclear source in the 
435$\mu$m continuum image of the galaxy obtained with ALMA led Xu et al.~(\cite{Xu15}) to rule out a Compton-thick AGN in \object{NGC~1614}.

The molecular gas in this source, imaged by the OVRO, SMA, CARMA, and ALMA interferometers,  is 
distributed in a nuclear disk and a fainter extension (Wilson et al.~\cite{Wil08}; Olsson et al.~\cite{Ols10}; K\"onig et al.~\cite{Kon13}; 
Sliwa et al.~\cite{Sli14}; Xu et al.~\cite{Xu15}; Usero et al.~in prep.). The recent $0\farcs5$--CO(2--1) 
map of the SMA, published by K\"onig et al.~(\cite{Kon13}), spatially resolved the nuclear disk into an asymmetric ring of $\sim230$~pc radius, similar to the CO(6--5) image of the central region obtained with ALMA by Xu et al.~(\cite{Xu15}). The disk shows an extension to the northwest in the SMA image. However, the lower resolution ($3\arcsec$) CO(1--0) map of CARMA (Sliwa et al.~\cite{Sli14}) shows an extension to the east.

The HST-NICMOS observations published by Alonso-Herrero et al.~(\cite{Alo01}) show a compact starburst nucleus (with an estimated age of $>10$~Myr) surrounded by a 
$\sim2\arcsec$ (600~pc) diameter ring of supergiant H{\small II} regions revealed by Pa${\alpha}$ line emission, which is associated with a younger starburst (5--10~Myr).  Lower level emission 
is also detected farther out to the east over a spiral structure out to $r\sim10\arcsec$ ($\sim3$~kpc). The luminosities of the giant H{\small II} regions in the ring are an order of magnitude brighter than 30Dor. The nuclear ring shows an extremely high value of the H${\alpha}$ surface brightness: $\sim60\times10^{41}$erg~s$^{-1}$kpc$^{-2}$ or, equivalently, a $SFR$ surface density of 
$\sim30~M_{\odot}$yr$^{-1}$kpc$^{-2}$. Such values are comparable to those derived across several kpc in high-z submillimeter galaxies (Men\'endez-Delmestre et al.~\cite{Men13}). The star-forming ring 
is also prominent in the  high-resolution radio continuum images (Olsson et al.~\cite{Ols10}; Herrero-Illana et al.~\cite{Her14}) and in the mid-infrared PAH emission maps of the galaxy (D{\'{\i}}az-Santos et 
al.~\cite{Dia08}; V\"ais\"ainen et al.~\cite{Vai12}).

The star formation properties of the molecular gas in \object{NGC~1614} stand out as {\em \emph{non-standard}} on several counts. Among the objects studied by Garc\'{\i}a-Burillo et al.~(\cite{Gar12}), \object{NGC~1614} shows the most extreme (high) 
star formation efficiency of the dense molecular gas traced by the 1--0 line of HCN 
($SFE_{\rm dense} \equiv L_{\rm FIR}/L'_{\rm HCN}=3800~L_{\odot}$/(K~km~s$^{-1}$~pc$^2$)). Furthermore, Xu et al.~(\cite{Xu15}) find a breakdown of the Kennicutt-Schmidt (KS) law on the linear scale of 100 pc in the nuclear ring, based on the gas column densities derived from the CO(6--5) map.

The kinematics of the ionized gas traced by the H$\alpha$ line, which was studied by Bellocchi et al.~(\cite{Bel12}), show significant departures from circular motions.   Bellocchi et al.~(\cite{Bel12}) find a main component associated with a rotating gas disk and a second (broader) component blueshifted by up to 300~km~s$^{-1}$ relative to the systemic velocity ($v_{\rm sys}$) of the galaxy. The latter was interpreted as the signature of a dusty outflow in the ionized gas. The blueshifted  emission then comes from the outflow lobe oriented toward us, and the far-side lobe remains undetected at optical wavelengths owing to extinction. The projections of the kinematic major axes  show that the rotating disk and the outflow axis differ by $\sim90^\circ$, an indication  that the outflow is oriented along the minor axis of the disk. The estimated outflow rate for the ionized gas is $\sim44M_{\odot}$ yr$^{-1}$ (Colina, private communication).

\subsection{\object{IRAS~17208-0014}}

\object{IRAS~17208-0014} ($D_{\rm A}=167$~Mpc, $D_{\rm L}=181$~Mpc; 1$\arcsec$= 810~pc) is a coalesced merger classified as an H{\small II} ULIRG ($L_{\rm IR}=2.4 \times 10^{12}~L_{\sun}$)  with the multiwavelength characteristics of an obscured nuclear starburst ($SFR \sim 240~M_{\odot}$ yr$^{-1}$) as the main driving source of its 
bolometric luminosity (Momjian et al.~\cite{Mom03}; Nardini et al.~\cite{Nar09, Nar10}; Rupke \& Veilleux~\cite{Rup13}; Teng \& Veilleux~\cite{Ten10}; Iwasawa et al. ~\cite{Iwa11}). However, the morphology of the X-ray source, analyzed by Iwasawa et al.~(\cite{Iwa11}),  shows hard X-ray emission and the tentative detection of a  high-ionization Fe K$\alpha$ line toward the NED nuclear position. Gonz\'alez-Mart\'{\i}n et al.~(\cite{Gon09}) classify the source as a candidate Compton-thick AGN with $L_{\rm X} \sim 10^{43}$erg~s$^{-1}$, leaving room for an obscured but not energetically dominant AGN in the nucleus of \object{IRAS~17208-0014}: $L_{\rm AGN}/L_{\rm bol} \sim 0.1\%$. However, Rupke \& Veilleux~(\cite{Rup13}) estimate this fraction to be $\sim11\%$ based on MIR diagnostics. The MIR 8.7$\mu$m continuum emission appears clearly extended over $\sim1.6$~kpc. The galaxy also shows extended 11.3$\mu$m PAH emission, but the equivalent width of this PAH feature indicates the presence of an additional nuclear continuum source that might be associated with the obscured AGN (Alonso-Herrero et al.~\cite{Alo14}).

The optical morphology of the galaxy is patchy because of dust obscuration,  but there is a well-defined nuclear source embedded in an extended disk of $\sim$2~kpc-radius detected at near infrared (NIR) wavelengths (Scoville et al.~\cite{Sco00}). 
The NIR disk has a $r^{1/4}$ elliptical-like profile. The outer disk of the galaxy shows two tidal tails, which betray the past interaction (Melnick \& Mirabel~\cite{Mel90}; Murphy et al.~\cite{Mur96}).  The more recent high-resolution AO-assisted NIR observations done by Keck (Medling et al.~\cite{Med14}) have resolved the inner disk into two overlapping nuclear disks with a small separation of  $\sim0\farcs2$ (200~pc).

Star formation is currently fueled by a massive molecular gas reservoir ($\sim (6-7) \times 10^{9}~M_{\sun}$: Downes \& Solomon~\cite{Dow98}; Wilson et al.~\cite{Wil08}).
The molecular disk of  \object{IRAS~17208-0014} was imaged in the CO(1--0) line with the PdBI with a $\sim3\farcs5$-spatial resolution by Downes \& Solomon~(\cite{Dow98}). The $2\arcsec$ (1.6~kpc)--diameter disk, barely resolved in these observations, shows a rotating 
pattern. The $\sim 1\arcsec$--CO(3--2) SMA map of Wilson et al.~(\cite{Wil08}) reveals a compact $\sim1\farcs5$~(1.2~kpc) diameter disk with 
fainter extensions farther out to the northwest and to the east.

Superposed on the overall rotating pattern, which dominates the kinematics of the disk, there is multiwavelength evidence of an outflow found in the ionized and in the neutral gas.   In spite of previous negative evidence of an outflow in the ionized gas (Westmoquette et al.~\cite{Wes12}; Rupke \& Veilleux~\cite{Rup13}), Arribas et al.~(\cite{Arr14}) report the detection of an outflowing component, based on very high S/N H$\alpha$ spectroscopy.  The maximum velocities of the ionized outflow reach moderate values of up to $v-v_{\rm sys}\sim-300$~km~s$^{-1}$ and the estimated outflow rate for the ionized gas is $\sim16~M_{\odot}$ yr$^{-1}$ (Colina, private communication). There is additional  evidence of an outflow in the neutral gas in  \object{IRAS~17208--0014}. 
Flows with maximum velocities 
of up to $v-v_{\rm sys}\sim-650$~kms$^{-1}$ and mass rates d$M$/d$t\sim 34~M_{\odot}$ yr$^{-1}$  were identified in the atomic gas in a spatially extended ($r\sim1.6$~kpc) structure traced by Na{\small I}~D absorption lines (Rupke \& Veilleux~\cite{Rup13}). Furthermore,  outflow velocities up to $v-v_{\rm sys}\sim-370$~km~s$^{-1}$ with associated outflow rates of 
$\sim 90~M_{\odot}$ yr$^{-1}$ were also identified in the cold molecular gas through OH absorption lines (Sturm et al.~\cite{Stu11}). However, this absorption line study, done with a spatial resolution of $\sim10\arcsec$, was not optimized to derive the size and the geometry of the molecular outflow structure, leaving the true outflow rate largely unconstrained. More recently, the optical and near-infrared emission line ratios analyzed by Medling et al.~(\cite{Med15}) have confirmed  that there is a multiphase outflow of shocked gas extending over a wide range of spatial scales from $r\sim400$~pc to $r\sim5$~kpc in \object{IRAS~17208--0014}.

\section{Observations}\label{Obs}
\subsection{NGC~1614}\label{1614-obs}

Observations of \object{NGC\,1614} were carried out with the PdBI array (Guilloteau et al.~\cite{Gui92}) between June 2012 and March 2013. 
We used the ABD configurations and six antennae. We observed the  $J$=1--0 line of CO (115.271\,GHz at rest).  During the observations we 
used the two polarizations of the receiver and the 3.6GHz-wide WideX spectral correlator of the PdBI; this is equivalent to 
9500\,km~s$^{-1}$ at the working frequency. Rest frequencies were corrected for the recession velocity initially assumed to be $v_{o}(HEL)=4795$\,km~s$^{-1}$.  The systemic velocity, redetermined in this work, is 
$v_{\rm sys}(HEL)=4763$\,km~s$^{-1}$.  Observations were conducted in 
single-pointing mode with a field of view (primary beam size) of  43$^{\prime\prime}$  centered at $\alpha_{2000}=04^{h}34^{m}00.03^{s}$ 
and $\delta_{2000}=-08^{\circ}34^{\prime}44.57\arcsec$. The latter is very close to  the nominal positions of the optical and near infrared centers, 
as determined by Neff et al.~(\cite{Nef90}) and Skrutskie et al.~(\cite{Skr06}). We redetermined the dynamical center of the galaxy in this work, 
which coincides within the errors with  the 
central peak detected in the 5~GHz radio continuum images obtained by Olsson et al.~(\cite{Ols10}) with MERLIN ($\alpha_{2000}=04^{h}
34^{m}00.03^{s}$ and $\delta_{2000}=-08^{\circ}34^{\prime}45.01\arcsec$). Visibilities were obtained through on-source integration 
times of 20 minutes framed by short ($\sim$2\,min) phase and amplitude calibrations on a nearby quasar.
The absolute flux scale in our maps was derived to 10$\%$ accuracy based on the observations of primary calibrators whose fluxes were 
determined from a combined set of measurements obtained at the 30m telescope and the PdBI array. The bandpass calibration is accurate to 
better than 5$\%$.

The image reconstruction was done with the
standard IRAM/GILDAS software (Guilloteau \& Lucas~\cite{Gui00}). We used both natural (NA) and uniform (UN) weighting to generate two different 
versions of the CO line map with a size of 51$\arcsec$ and $0\farcs1$/pixel sampling. The corresponding synthesized beams are 
$3\farcs8\times1\farcs2$ at  $PA=12^{\circ}$ for the NA weighting map, and $1\farcs5\times0\farcs8$ at  $PA=23^{\circ}$ for the UN 
weighting map. The conversion factors between flux (Jy\,beam$^{-1}$) and temperature units (K) are  20\,K~Jy$^{-1}$~beam and 82\,K~Jy$^{-1}$~beam for the NA and 
UN weighting datasets, respectively. The corresponding point source sensitivities derived from emission-free channels are 0.8\,mJy~beam$^{-1}$ (NA)  and 0.9\,mJy~beam$^{-1}$ (UN) in 9.8~MHz (26~km~s$^{-1}$)-wide channels.  The actual noise levels increase by up to a factor 2--3 in channels 
with strong line-emission due to inherent limitations in the deconvolution process. An image of the continuum emission of the galaxy was obtained 
by averaging those channels free of line emission. The point source sensitivity for the continuum map is 0.1\,mJy~beam$^{-1}$ in the highest 
resolution version of the data obtained by adopting UN weighting.

 \begin{figure*}[tbh!]
   \centering  
  \includegraphics[width=15cm]{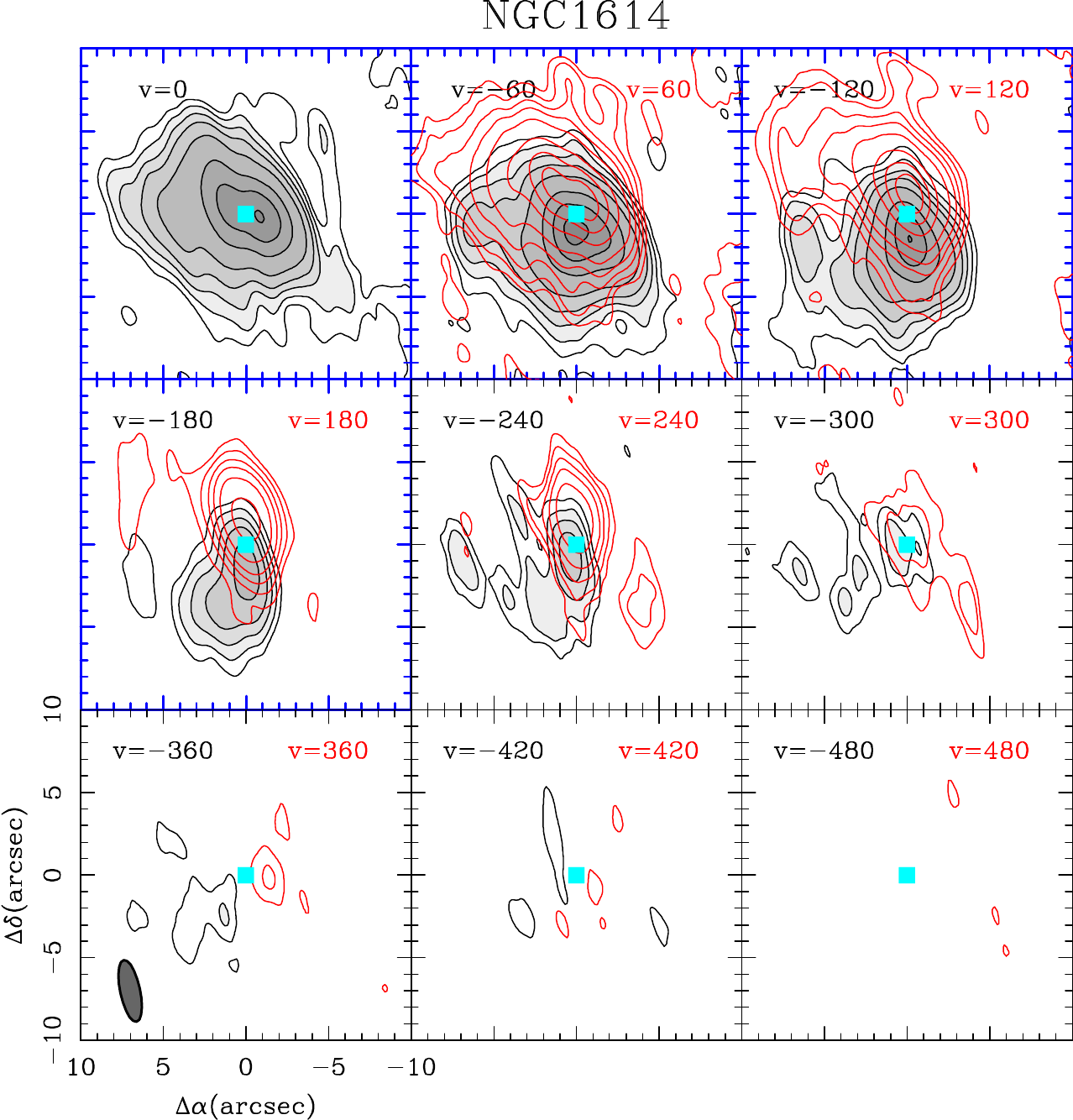} 
       \caption{CO(1--0) velocity-channel maps observed with the PdBI in the nucleus of \object{NGC~1614} with a spatial resolution of  
$3\farcs8\times1\farcs2$ at  $PA=12^{\circ}$ obtained with natural weighting (beam is plotted as a filled ellipse in the bottom left corner of the lower left panel). Velocity resolution is 60~km~s$^{-1}$. We show a field of view of 20$\arcsec$, i.e., $\sim$1/2 the diameter of the primary beam at 113.5~GHz.  We display channel maps grouped by pairs from 
$v-v_{\rm sys}=0$~km~s$^{-1}$ in steps of 60~km~s$^{-1}$ with  $v_{\rm sys}$(HEL)$~=4763$~km~s$^{-1}$ from --480~km~s$^{-1}$ to 480~km~s$^{-1}$. Emission at blueshifted velocities is displayed in gray scale and black contours, and emission at redshifted velocities is displayed in red contours. Contour levels are 3$\sigma$, 5$\sigma$, 8$\sigma$, 12$\sigma$, and  20$\sigma$ with 1$\sigma=0.5$~mJy~beam$^{-1}$ for the channels of the line wings ($\mid$$v-v_{\rm sys}$$\mid$$~\geq 210$~km~s$^{-1}$), and  3$\sigma$, 5$\sigma$, 8$\sigma$, 12$\sigma$, 20$\sigma$ to 80$\sigma$ in steps of  20$\sigma$, with 1$\sigma=1.5$~mJy~beam$^{-1}$ for the channels of the line core ($\mid$$v-v_{\rm sys}$$\mid$$~< 210$~km~s$^{-1}$; boxes highlighted in blue). The position of the dynamical center ([$\Delta\alpha$, $\Delta\delta$]~=~[0$\arcsec$,0$\arcsec$]~=~[$\alpha_{2000}=04^{h}34^{m}00.03^{s}$, $\delta_{2000}~=-08^{\circ}34^{\prime}45.01\arcsec$]) is highlighted by the (blue) square marker.}
              \label{NGC1614-channels}
\end{figure*}
   

 \begin{figure*}[tbh!]
   \centering  
  \includegraphics[width=17cm]{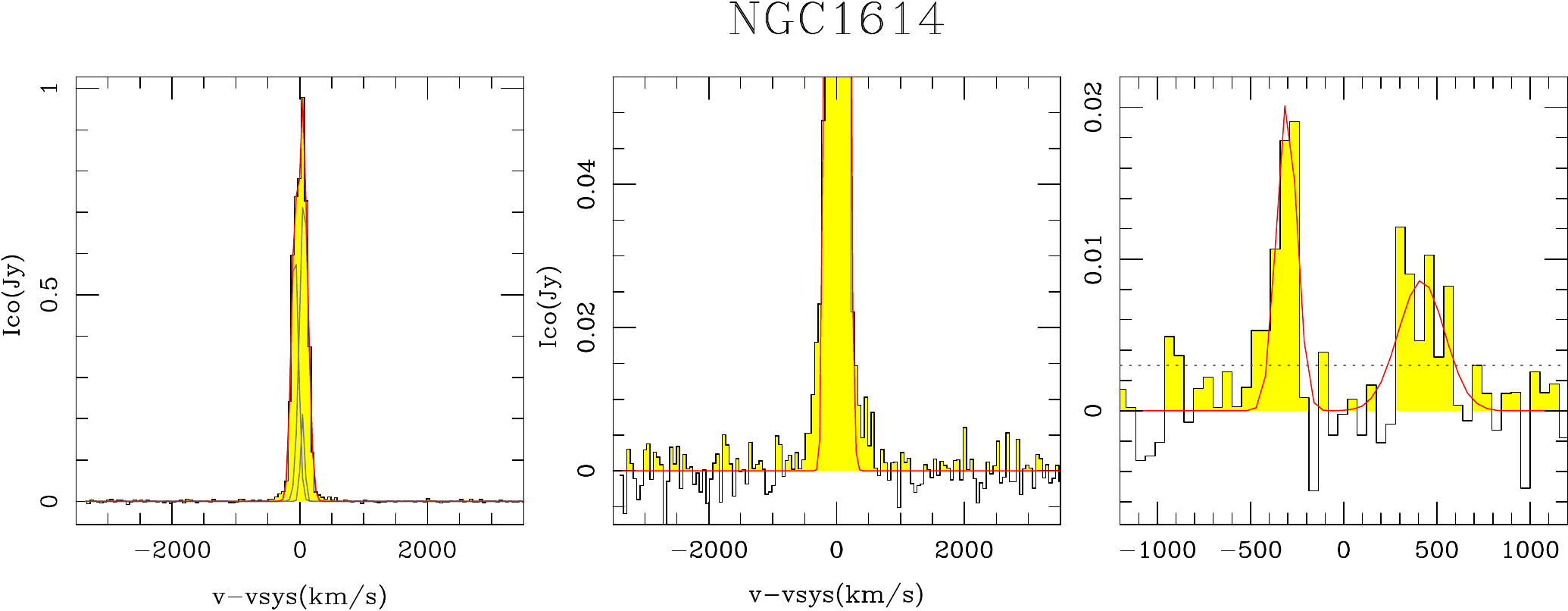}
    
       \caption{CO(1--0) spectrum obtained by spatially integrating  the emission over the disk of \object{NGC~1614}. We show the three individual Gaussian
       components fitted to the spectrum (gray curves) and the implied overall fit combining them  (red curve) ({\it left panel}). A close-up view of the 
       overall fit to the spectrum  is shown in the {\it middle panel} to highlight emission from the line wings. Residuals and associated Gaussian fits are shown in 
       the {\it right panel} with the typical 1$\sigma$ level indicated by the dotted line.}
              \label{spectra-1614}
\end{figure*}
   

\subsection{IRAS~17208-0014}\label{17208-obs}

Observations of \object{IRAS~17208-0014} were carried out with the PdBI in March 2012. 
We used the B configuration and six antennae of the array. We observed the  $J$=2--1 line of CO (230.538\,GHz at rest).  As for 
\object{NGC~1614}, we used  the two polarizations of the receiver and the WideX correlator, and this is equivalent to 
4900\,km~s$^{-1}$ at the CO(2--1) frequency. Rest frequencies were corrected for an adopted recession velocity of $v_{o}(HEL)=12838$\,km~s$^{-1}$.  We derived the systemic velocity in this work, which is  $\sim$30\, km~s$^{-1}$ blueshifted relative to $v_{o}$: $v_{\rm sys}(HEL)=12808$\,km~s$^{-1}$.  We used a single-pointed observing mode with a primary beam size of  23$^{\prime\prime}$  centered at  $\alpha_{2000}=17^{h}23^{m}21.90^{s}$  and $\delta_{2000}=-00^{\circ}17^{\prime}00.90\arcsec$, which is $<1\arcsec$ away from the radiocontiuum peak detected in
the 1.4\,GHz VLBI maps of Momjian et al.~(\cite{Mom03}) and the dynamical center of the galaxy, which was determined in this work ($\alpha_{2000}=17^{h}23^{m}21.96^{s}$  and $\delta_{2000}=-00^{\circ}17^{\prime}00.87\arcsec$). 
 Visibilities were obtained through on-source integration  times of 20 minutes framed by short ($\sim$2\,min) phase and amplitude calibrations on  
 a nearby quasar. The absolute flux scale in our maps was derived to 10$\%$ accuracy, and the bandpass calibration is accurate to better than 5$\%$.

We performed image deconvolution using IRAM/GILDAS software. As for \object{NGC~1614}, we used both NA and UN weighting to generate two versions of the CO map with a size of 26$\arcsec$ and $0\farcs1$/pixel sampling. The corresponding synthesized beams are $0\farcs8\times0\farcs5$ at  $PA=18^{\circ}$ for the NA weighting map, and $0\farcs6\times0\farcs5$ at  $PA=61^{\circ}$ for the UN weighting map. The conversion factors between flux (Jy\,beam$^{-1}$)  and temperature units (K) are  62\,K~Jy$^{-1}$~beam and 84\,K~Jy$^{-1}$~beam for the NA and UN weighting datasets, respectively. The corresponding point source sensitivities derived from emission-free channels are 0.9\,mJy~beam$^{-1}$ (NA)  and 1.2\,mJy~beam$^{-1}$ (UN) in 20~MHz (27~km~s$^{-1}$)-wide channels.  The noise figures are to be increased by a factor 2--3 in channels with strong line emission. A continuum image of the galaxy was obtained by averaging channels free of line emission with a point-source sensitivity of 0.1\,mJy~beam$^{-1}$ in the map derived with UN weighting.

 \begin{figure*}[tbh!]
   \centering  
  \includegraphics[width=17cm]{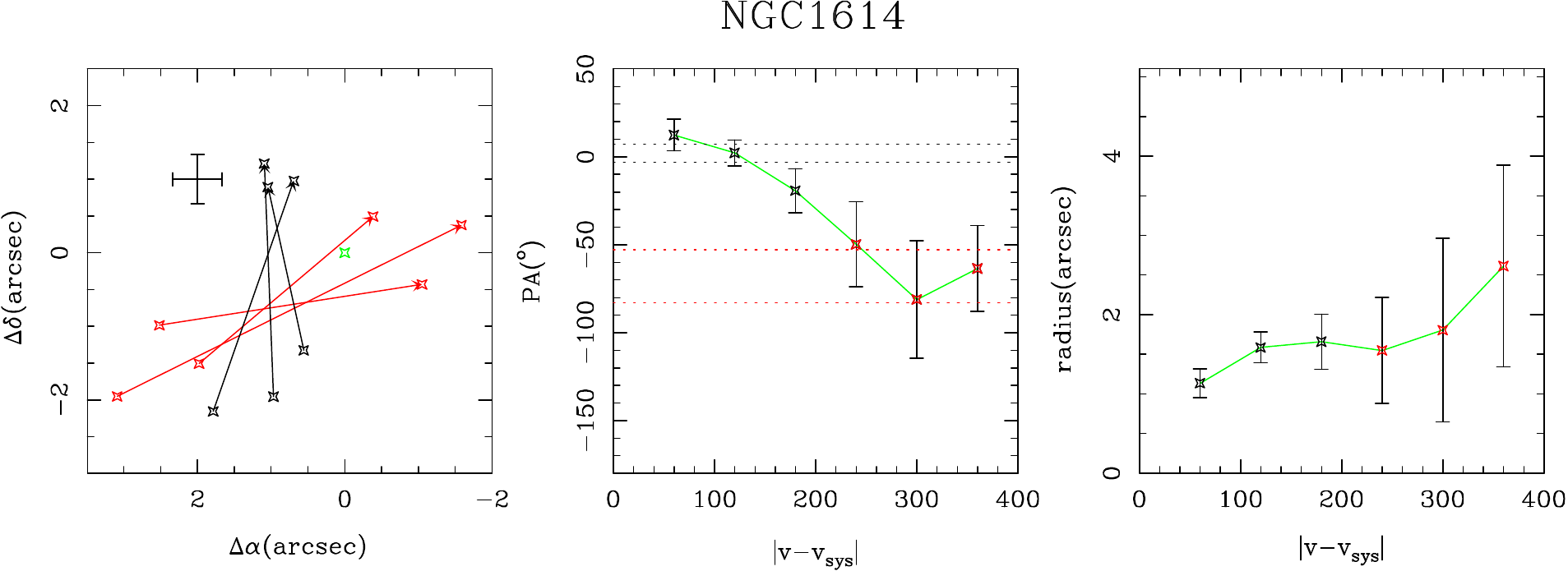}
  
       \caption{
       Centroids  of CO(1--0) emission in \object{NGC~1614} (in [$\Delta\alpha$, $\Delta\delta$]-units) derived 
       with a velocity resolution of 60~km~s$^{-1}$, using a power law of the flux as weighting function with index $n=1/2$, are displayed in the {\it left panel} . Emission centroids calculated for velocity channels that are equidistant from $v_{\rm sys}$ (i.e., with a  common $\mid$$v-v_{\rm sys}$$\mid$--offset) are connected to visualize the change of $PA$, defined for the redshifted component of the connected centroids and measured east from north, as a function of velocity  ({\it middle panel}). We also plot the change in the average radius $r$, defined as half of the mutual distance from the connected centroids,  as a function of velocity ({\it right panel}). Red color is used  for velocity channels  of the  line wing and black color for the line core. The dotted black (red) line in the {\it middle panel} identifies the $\pm1\sigma$ range around the average $PA$ of the disk as derived inside the line core (line wing): $PA_{\rm core}=2\pm5^{\circ}$ ($PA_{\rm wing}=-68\pm15^{\circ}$). The green marker in the {\it left panel} identifies the dynamical center. Error bars as a function of velocity channel are shown in the {\it middle} and {\it right} panels. Error bars in the {\it left panel} are for the line core.}
              \label{PA-dist-1614}
\end{figure*}
   

\section{Results}\label{Results}
\subsection{\object{NGC~1614}: channel maps and emission centroids}\label{1614-ch}

Figure~\ref{NGC1614-channels} shows the CO(1--0) velocity-channel maps observed in \object{NGC~1614} with a spatial 
resolution of  $3\farcs8\times1\farcs2$ at  $PA=12^{\circ}$, obtained using natural weighting of the data to recover the maximum percentage of low-level emission in extended structures.  We display channel maps grouped by pairs with a velocity spacing of 60~km~s$^{-1}$ from $v-v_{\rm sys}=0$~km~s$^{-1}$ with  
$v_{\rm sys}$(HEL)$~=4763$~km~s$^{-1}$, a value determined by the best-fit kinemetry solution found for the mean velocity field described in Sect.~\ref{1614-mom}.

Virtually all the CO line emission detected over significant 3$\sigma$-levels comes from the central $r \sim 10\arcsec$ (3~kpc) region that is 
displayed in Fig.~\ref{NGC1614-channels}.  Overall, this emission extends over a roughly symmetric $\sim900$~km~s$^{-1}$-wide velocity span: $v-v_{\rm sys} \sim [+450,-450]$~km~s$^{-1}$. However, the bulk of the emission, up to $\simeq 98\%$ of the grand total, comes from a more restricted  velocity range: $\mid v-v_{\rm sys} \mid \leq 210$~km~s$^{-1}$, hereafter referred to as the line core.

Figure~\ref{spectra-1614} shows the CO spectrum spatially integrated over a square region of 4~kpc size to cover the full extent of the \object{NGC~1614} disk. 
The emission from the line core can be satisfactorily fit by three Gaussian components. In this fit we assigned three initial values for the velocity centroids of the Gaussian components: one around $v_{\rm sys}$, and two velocities symmetrically offset by $\pm100$~km~s$^{-1}$ with respect to $v_{\rm sys}$ to cover the expected symmetric range  caused by rotation found in the model of Bellocchi et al.~(\cite{Bel12}). However, outside the line core 
there is an excess of fainter but nevertheless statistically significant emission  (up to $\sim11\sigma$ in velocity-integrated units) which extends up to $\mid v-v_{\rm sys} \mid \sim 450$~km~s$^{-1}$,  
hereafter referred to as the line wings (i.e., $210$~km~s$^{-1}$$< \mid v-v_{\rm sys} \mid \leq 450$~km~s$^{-1}$).  Emission stemming from the line core shows the signature of a rotating disk with a kinematic 
major axis oriented roughly north-south: emission at redshifted (blueshifted) velocities comes from the northern (southern) side of the disk, as can 
be guessed by inspecting Fig.~\ref{NGC1614-channels}. By contrast, outside the line core, the kinematic major axis for the line wings 
seems to tilt progressively toward the east-west axis (see Fig.~\ref{NGC1614-channels}).

 \begin{figure*}[tbh!]
   \centering  
  \includegraphics[width=16cm]{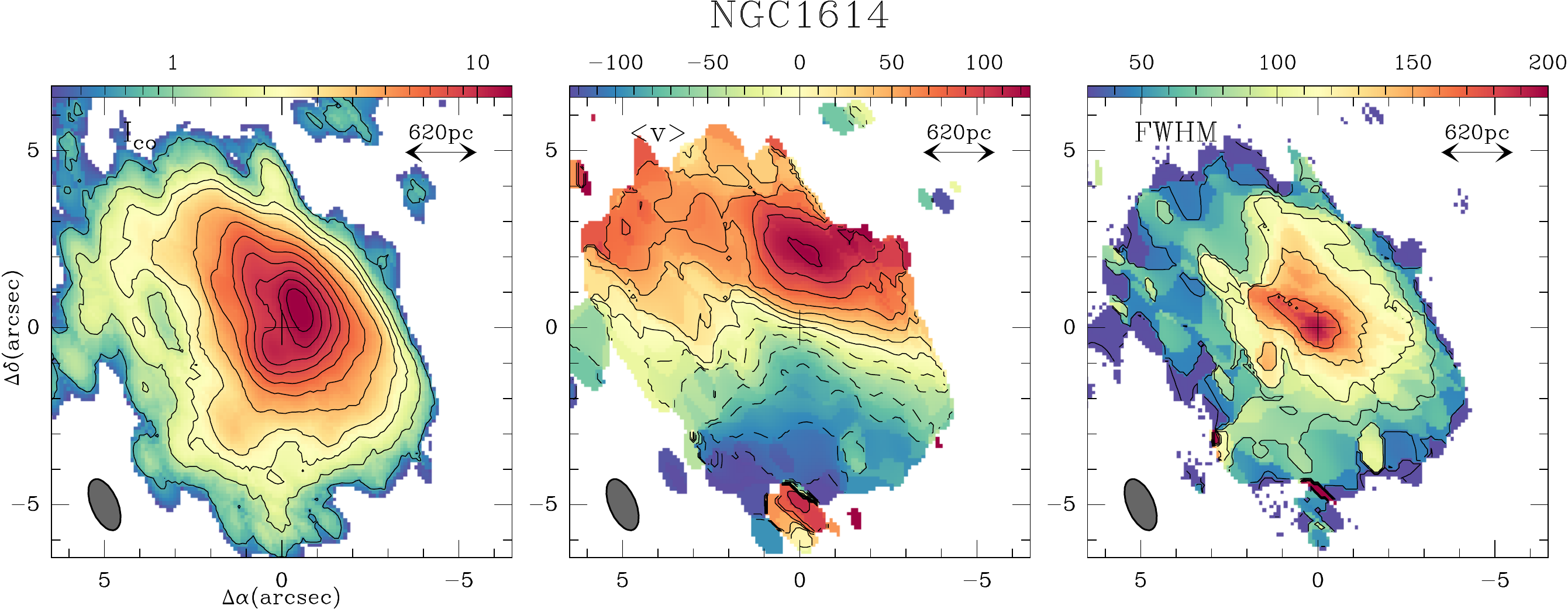}
  \caption{{\it Left panel}:~CO(1--0) integrated intensity map obtained in the disk of \object{NGC~1614}. The map is shown on color scale with contour levels  5$\%$, 10$\%$, 15$\%$, 20$\%,$ to  90$\%$ in steps of 10$\%$ of the peak value~=14~Jy~km~s$^{-1}$beam$^{-1}$ ($\sim50\sigma$).  
{\it Middle panel}:~The CO(1--0) isovelocity contours spanning the range (--120~km~s$^{-1}$, 120~km~s$^{-1}$) in steps of 20~km~s$^{-1}$ are overlaid on a false-color velocity map (linear color scale as shown). Velocities refer to $v_{\rm sys}$(HEL)$~=4763$~km~s$^{-1}$. {\it Right panel}:~Overlay of the CO(1--0) line widths (FWHM) shown in contours (30, 45 to 195~km~s$^{-1}$ in steps of 30~km~s$^{-1}$) on a false-color width map (linear scale as shown). The filled
ellipses in the bottom left corner of each panel represent the CO(1-0) beam size obtained with uniform weighting ($1\farcs5\times0\farcs8$ at  $PA=23^{\circ}$). The position of the dynamical center  is highlighted by the cross marker.}
              \label{NGC1614-moments}
\end{figure*}
   

 \begin{figure*}[tbh!]
   \centering  
  \includegraphics[width=16cm]{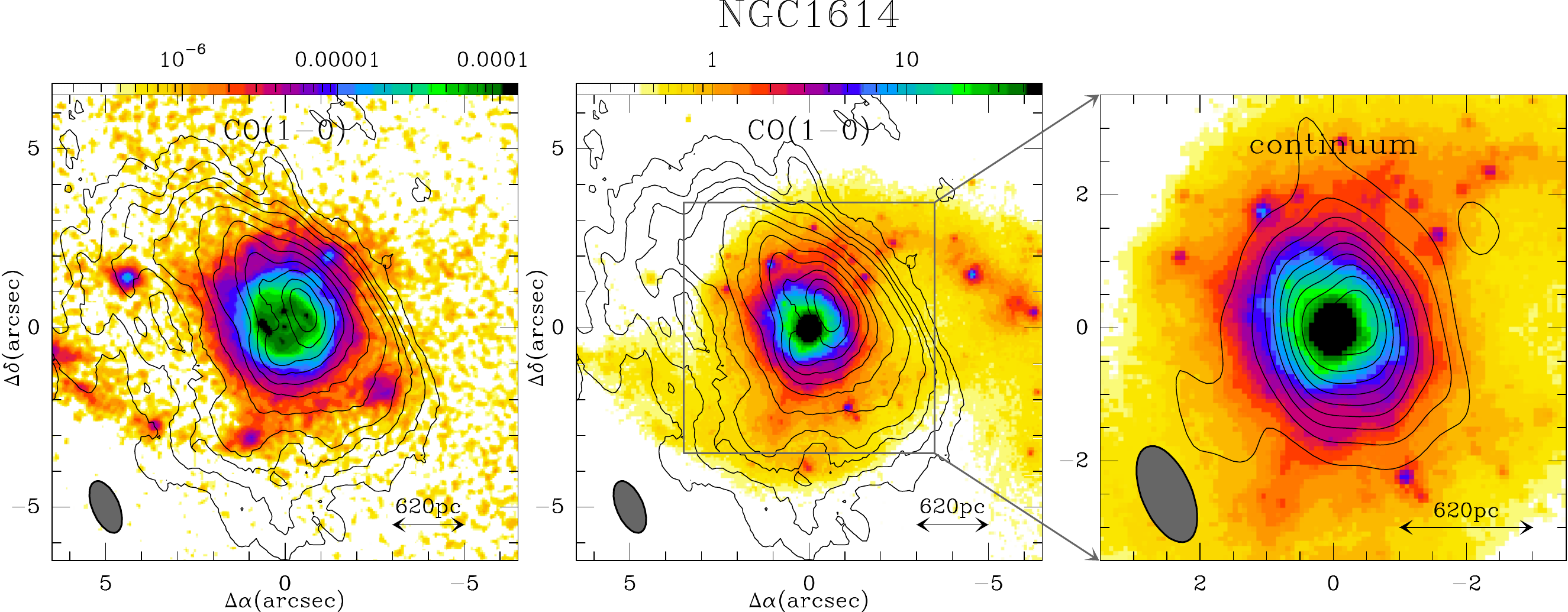}
  \caption{{\it Left panel}:~Overlay of the CO(1--0) intensity contours (levels as in the left-hand panel of Fig.~\ref{NGC1614-moments}) on the Pa$\alpha$ 
  emission HST map of \object{NGC~1614}  (color scale as shown in arbitrary units). {\it Middle panel}:~Same as {\it left panel} but with CO 
  contours overlaid on the HST/NICMOS 1.6$\mu$m continuum image  (color scale as shown in arbitrary units). {\it Right 
  panel}:~Same as {\it middle panel} but zooming in on the continuum emission image of \object{NGC~1614} at 
  113.5~GHz, in contours (10$\%$ to  90$\%$  in steps of 10$\%$ of the peak value~=~2.4~mJy~beam$^{-1}\sim24\sigma$), overlaid on the HST 
  NICMOS image. The filled ellipses represent the  CO beam size.}
              \label{NGC1614-overlay}
\end{figure*}
   

In an attempt to quantify the kinematic decoupling of the emission of the line wings relative to the  line core, we have derived the 
centroids  of CO emission in \object{NGC~1614} in [$\Delta\alpha$, $\Delta\delta$]-units. The left-hand panel of Fig.~\ref{PA-dist-1614} shows the CO centroids for 
the same velocity channels displayed in Fig.~\ref{NGC1614-channels}. 
The centroids  of CO(1--0) emission are derived using as weighting function a power law of the flux measured at a given offset within the inner $20\arcsec$ field of view with an index $n=1/2$ in order to optimize the contribution from extended emission and reduce accordingly the noise caused by clumpiness in this determination. Emission centroids calculated for velocity channels which are equidistant from $v_{\rm sys}$ are connected so as to visualize the change in $PA$ as a function of velocity, shown in the middle panel of Fig.~\ref{PA-dist-1614}. The  $PA$ is defined for the redshifted component of the connected centroids, and it is measured east from north (i.e., negative values to the west). The right-hand panel of Fig.~\ref{PA-dist-1614} illustrates  the change in average radius $r$, defined as half of the mutual distance between the connected centroids, as a function of velocity. The left-hand panel of Fig.~\ref{PA-dist-1614} identifies  the dynamical center at [$\Delta\alpha$, $\Delta\delta$]=[$0\arcsec,0\arcsec$] as determined in Sect.~\ref{1614-mom}. The centroid barycenter is slightly shifted relative to the dynamical center because of the non-axisymmetries in the brightness distribution of CO.

 As illustrated in the middle panel of Fig.~\ref{PA-dist-1614}, the value derived for $PA$ changes by $\sim 80^\circ$ from  $\mid v-v_{\rm sys} \mid$~=~60~km~s$^{-1}$, where $PA\sim+13^\circ$,  to 
$\mid v-v_{\rm sys}\mid$~=~360~km~s$^{-1}$, where $PA\sim -64^\circ$. We estimate a weighted mean for $PA$ in the line core of 
$PA_{\rm core}=2\pm5^{\circ}$, and this is close to the value derived for $PA$ using the kinemetry fit  of Sect.~\ref{1614-mod} ($PA\sim -8^\circ\pm 5^\circ$). The corresponding weighted mean for $PA$ in the line wing region is $PA_{\rm wing}=-68\pm15^{\circ}$. We can therefore conclude that the evidence of
a kinematic decoupling, reflected in the measured difference $PA_{\rm wing}-PA_{\rm core}\sim-70\pm16^{\circ}$ is a significant $\simeq4.4\sigma$ result. 
However, the right-hand panel of Fig.~\ref{PA-dist-1614} shows that there is no statistically significant change in the radial distance, which stays fairly constant at a value of $r\sim1\farcs5-2\arcsec$~($0.5-0.6$~kpc) as a function of the velocity offset.

 \begin{figure*}[tbh!]
   \centering  
  \includegraphics[width=17cm, angle=0]{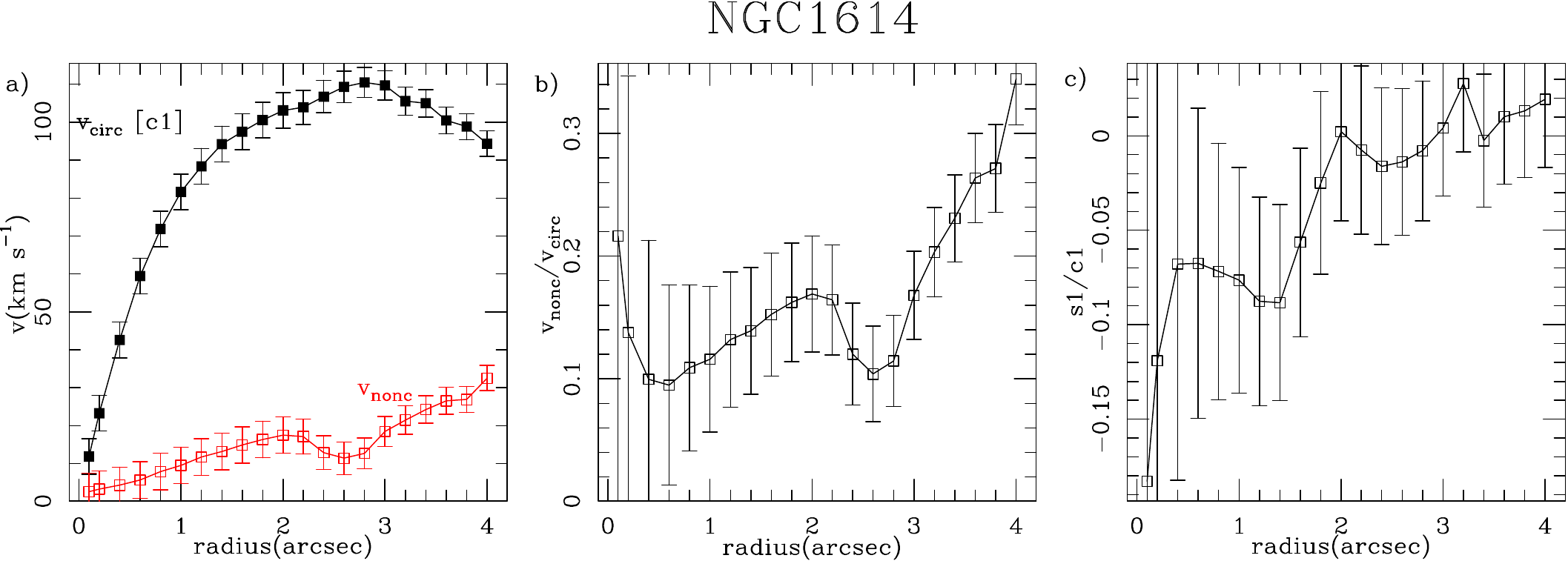}
  
       \caption{{\bf a)}~ $c_1$ radial profile (black curve) derived from the Fourier decomposition of the velocity field of \object{NGC~1614}. The  $c_1$ term accounts for the (projected) circular component of the velocity field ($v_{\rm circ}$).  We also plot (red curve) the radial profile of the (projected) non-circular motions ($v_{\rm nonc}$) derived up to the third order of the decomposition. {\bf b)}~The radial profile of the $v_{\rm nonc}/v_{\rm circ}$ ratio. {\bf c)}~The radial variation of the  $s_{1}/c_{1}$. The $s_{1}$ term represents the (projected) axisymmetric radial motions of the velocity field. Errorbars in all panels represent 1$\sigma$ errors.} 
              \label{1614-kinemetry}
\end{figure*}
   

 \begin{figure*}[tbh!]
   \centering  
  \includegraphics[width=11.5cm, angle=0]{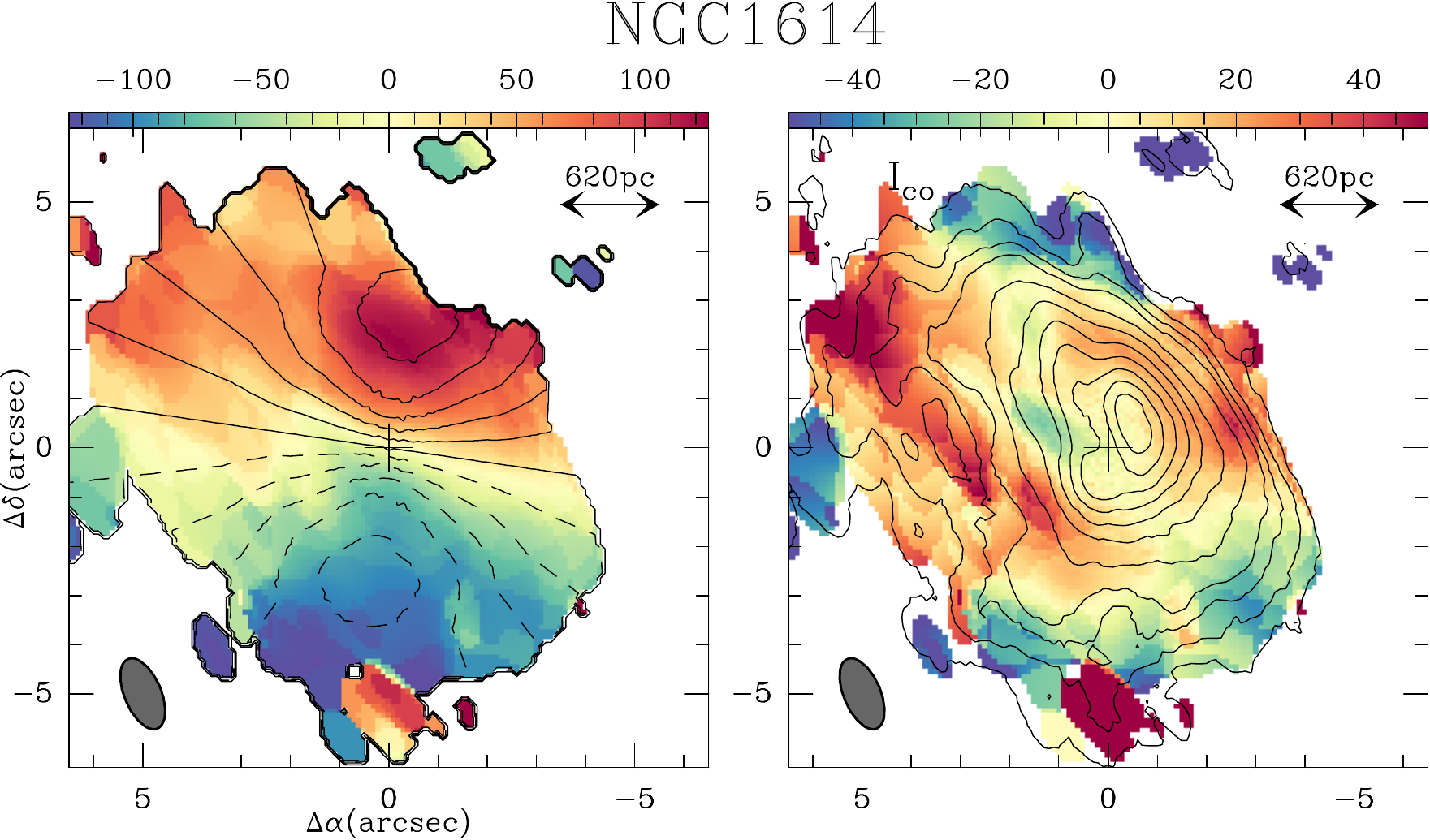}
     
       \caption{{\it Left panel}:~Overlay of the velocity field of the best-fit rotating disk model (in contours spanning the range --100~km~s$^{-1}$ to 100~km~s$^{-1}$ in steps of 20~km~s$^{-1}$) on the CO(1--0) isovelocities (color scale as shown) of \object{NGC~1614}. {\it Right panel}:~Overlay of the  CO(1--0) intensity contours (levels as in the left-hand panel of Fig.~\ref{NGC1614-moments}) on the residual mean-velocity field (in color scale as shown) obtained
after subtraction of the best-fit rotating disk model from the observations.}
              \label{1614-kinemetry-residuals}
\end{figure*}
   

\subsection{\object{NGC~1614}: moment maps}\label{1614-mom}

Figure~\ref{NGC1614-moments} shows the first three moment maps derived from the CO(1--0) line data obtained by applying uniform weighting to the visibilities inside the velocity interval  $v-v_{\rm sys}$~=~[+600,--600]~km~s$^{-1}$. Uniform weighting allows us to reach a spatial resolution 
of $1\farcs5\times0\farcs8$ at  $PA=23^{\circ}$. The velocity-integrated intensity map shown in the left-hand panel of Fig.~\ref{NGC1614-moments}  was derived assuming a $3\sigma$-clipping on the intensities. The mean-velocity field and the velocity width maps, shown in the middle and right-hand panels of  Fig.~\ref{NGC1614-moments}, respectively, were both derived by assuming a more demanding  $5\sigma$-clipping on the intensities to maximize the reliability of the images.

The left-hand panel of Fig.~\ref{NGC1614-moments} shows that the CO emission comes from a spatially-resolved  $11\arcsec \times 8\arcsec$ (3.4~kpc~$\times$~2.5~kpc) lopsided molecular disk. The disk shows an arc-like feature that runs from the northeast to the southeastern side. The strongest emission peak is located $\sim300$~pc northwest relative to the central ($0\arcsec,0\arcsec$)--offset.

The middle panel of Fig.~\ref{NGC1614-moments} confirms the signature of a spatially resolved rotating disk, as described in Sect.~\ref{1614-ch}. When assuming that the spiral features seen in the optical and NIR images of the galaxy are trailing, the gas rotation should be counterclockwise to account for the observed velocity gradient. This constrains the geometry of the galaxy disk with the eastern (western) side being the far (near) side.  The choice of the ($0\arcsec,0\arcsec$)--offset as the nucleus  maximizes the symmetry of the mean-velocity field. This location corresponds to the maximum velocity witdh (FWHM) measured in the disk, shown in the right-hand panel of Fig.~\ref{NGC1614-moments}, and also to the radio continuum peak of Olsson et al.~(\cite{Ols10}). Therefore this position qualifies as the dynamical center of the galaxy.  The molecular gas is very turbulent: we estimate an average FWHM$\sim$80~km~s$^{-1}$ over the disk. The value of FWHM is lowered to $\sim$70~km~s$^{-1}$ if we exclude the central $r=1.5\arcsec$~(0.5~kpc) of the disk to minimize the effects of beam smearing for our current spatial resolution. This implies an intrinsic velocity dispersion of $\sigma \simeq 30$~km~s$^{-1}$ on average.

An inspection of the middle panel of Fig.~\ref{NGC1614-moments} indicates that while the kinematic major axis is oriented roughly north-south, the kinematic minor axis is not oriented east-west, but rather at an angle ($\geq15^{\circ}$). This global tilt proves that there are significant departures from circular rotation. Additional wiggles in the velocity field, identified  on the eastern side of the disk, betray non-circular motions on smaller spatial scales.

The left-hand and middle panels of Fig.~\ref{NGC1614-overlay} overlay the CO line intensities on the HST  NICMOS images of \object{NGC~1614} obtained in the Pa$\alpha$ 
line and in the 1.6$\mu$m continuum by Alonso-Herrero et al.~(\cite{Alo01}). While a sizable fraction of the CO emission originates in the nuclear star-forming ring, which is barely resolved by the PdBI, fainter CO emission is also detected extending farther out to the east up to $r\sim6\arcsec$~(1.9~kpc), where strong 
Pa$\alpha$ emission identifies the star-forming complexes of the eastern spiral arm. The right-hand panel of Fig.~\ref{NGC1614-overlay} shows that, unlike the CO line emission, the mm-continuum emission at 113.5~GHz is  restricted to the nuclear ring.

 \begin{figure*}[tbh!]
   \centering  
  \includegraphics[width=17cm]{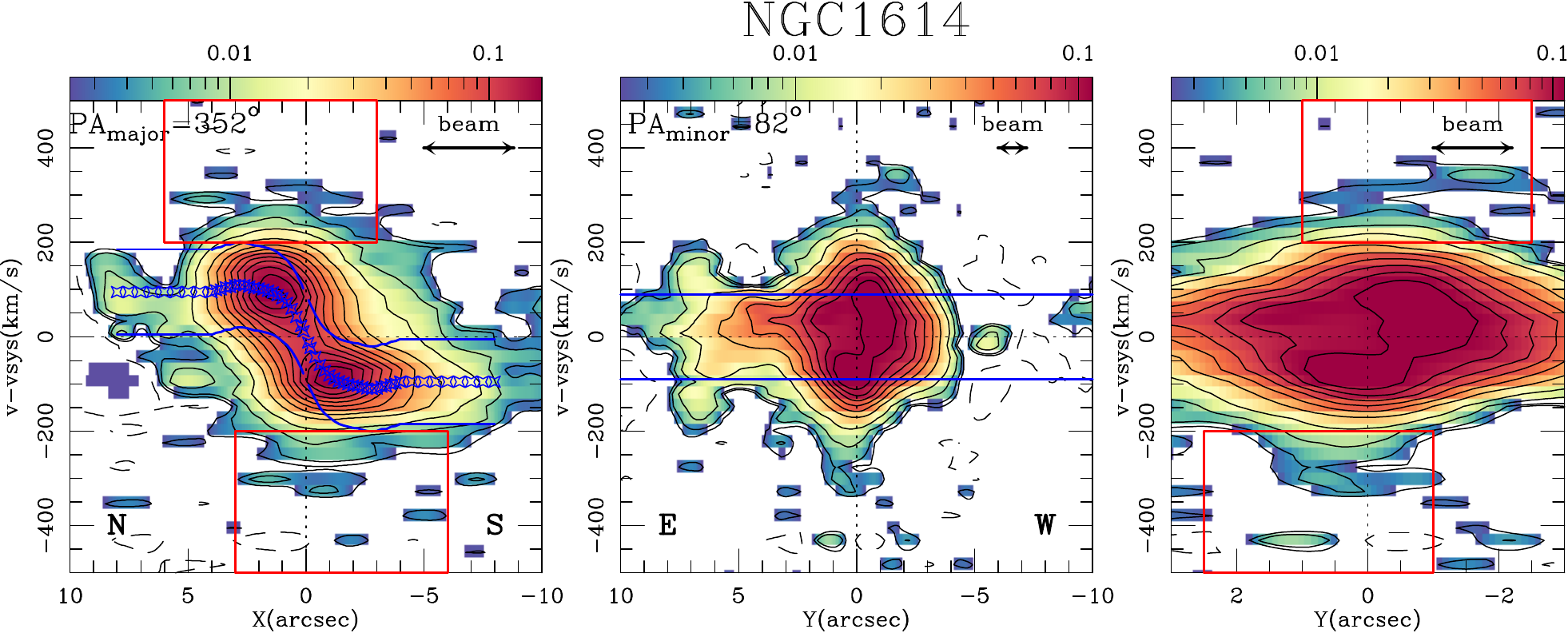}
       \caption{{\it Left panel:}~Position-velocity (p-v) plot taken along the kinematic major axis ($PA=352^{\circ}$) of \object{NGC~1614}. The blue star markers delineate  the (projected) rotation curve fitted in Sect.~\ref{1614-mod} and the allowed virial range around it (delimited by the blue lines). The red squares delimit the region where emission is detected from the line wings.  Contour levels are -3$\sigma$, 3$\sigma$, 5$\sigma$, 10$\sigma$, 20$\sigma,$ to 200$\sigma$ in steps of 20$\sigma$ with 1$\sigma=0.8$~mJy~beam$^{-1}$.            
   {\it Middle panel:}~The p-v plot taken along the kinematic minor axis ($PA=82^{\circ}$). Contour levels are -3$\sigma$, 3$\sigma$, 5$\sigma$, 10$\sigma$, 20$\sigma,$ to 140$\sigma$ in steps of 20$\sigma$ with 1$\sigma=0.8$~mJy~beam$^{-1}$. The allowed virial range is delimited by the blue curves.  {\it Right panel:}~Same as
  {\it middle panel} but zooming in on the central $\pm3\arcsec$ region around the center. The red squares delimit the region where emission is detected from the line wings. In all panels the velocity scales are relative to $v_{\rm sys}$, and the spatial 
scales (X and Y) are relative to the dynamical center.  We show the spatial resolutions projected along the major axis ({\it left panel}) and along the minor axis ({\it middle} and {\it right panels}) obtained with NA weighting.}
              \label{1614-slices}
\end{figure*}
   

\subsection{\object{NGC~1614}: kinematic modeling}\label{1614-mod}
 
\subsubsection{{\tt Kinemetry fit}: the coplanar solution}\label{1614-kin} 

To quantify the deviations from circular motions discussed in Sect.~\ref{1614-mom}, we used the mean-velocity field of Fig.~\ref{NGC1614-moments} to find  the best-fit Fourier decomposition of the line-of-sight velocities using the software package {\tt kinemetry}  (Krajnovi{\'c} et al.~\cite{Kra06}).  A key implicit assumption of this decomposition, which we question in Sect.~\ref{1614-outflow}, is that the gas kinematics in \object{NGC~1614} can be modeled by orbits that lie at all radii in a coplanar geometry.   

Figure~\ref{1614-kinemetry}, which shows the radial profiles of the main terms of the Fourier decomposition of the velocity field,  summarizes the results of this analysis. The  $c_1$ term, which accounts for the (projected) circular component of the velocity field ($v_{\rm circ}$), is the dominant contributor to the observed motions, as illustrated in Fig.~\ref{1614-kinemetry}a. Figure~\ref{1614-kinemetry-residuals} shows the rotating disk solution of {\tt kinemetry} and the  associated residuals. The geometry of the rotating disk is defined by a position angle $PA=-8\pm5\equiv352\pm5^{\circ}$ (measured east from north for the receding side of the major axis) and an inclination $i=33\pm2^{\circ}$. We also determined $v_{\rm sys}$(HEL)$~=4763\pm10$~km~s$^{-1}$, as part the best-fit solution. Relative to $v_{\rm circ}$, non-circular motions ($v_{\rm nonc}$) derived up to the third order of the Fourier decomposition, are within a significant 10\%\ to 30$\%$ range, as shown in Fig.~\ref{1614-kinemetry}b. Furthermore, Fig.~\ref{1614-kinemetry}c shows that the sign of the $s_{1}$ term, which represents the (projected) axisymmetric radial motions of the velocity field, is negative up to $r \sim 2\arcsec$~(0.6~kpc). This is indicative of inflowing motions at these radii for the adopted geometry of the disk when assuming the gas is coplanar. Deprojected, these in-plane motions would translate into a maximum amplitude of the radially inward field of $\leq$15~km~s$^{-1}$. These low amplitude inward motions can be explained by density-wave-driven gas flows.

 \subsubsection{A non-coplanar outflow solution}\label{1614-outflow}

 While the coplanar solution discussed in Sect.~\ref{1614-kin} could account for the observed trend of the {\em \emph{mean velocities}} fitted by {\tt kinemetry}, it nevertheless fails to explain the order of magnitude and, also,  the 2D pattern of the {\em \emph{high velocity}} emission found in the line wings, as argued in the following. 

The left-hand panel of Fig.~\ref{1614-slices} shows the position-velocity (p-v) plot taken along the kinematic major axis of the rotating disk ($PA=352^{\circ}$) with the best-fit (projected) rotation curve superposed. Most of the emission in the line core lies within the estimated {\it \emph{virial}} range. The latter is defined by a combination of circular rotation ($\sim \pm100$~km~s$^{-1}$), turbulence (FWHM/2$\sim \pm40$~km~s$^{-1}$), and an estimated upper limit to the contribution from in-plane non-circular motions of $\sim1/2~\times~v_{\rm circ}$\footnote{This amounts to $\pm 50$~km~s$^{-1}$ in \object{NGC~1614}.} (e.g., Colombo et al.~\cite{Col14}; Garc\'{\i}a-Burillo et al.~\cite{Gar14}). However, emission from the line wings, identified in Fig.~\ref{1614-slices}, goes beyond the adopted virial range, arguing in favor either of a fast(er)-rotating disk or of a non-rotationally supported component.  

 \begin{figure*}[tbh!]
   \centering  
  \includegraphics[width=18cm]{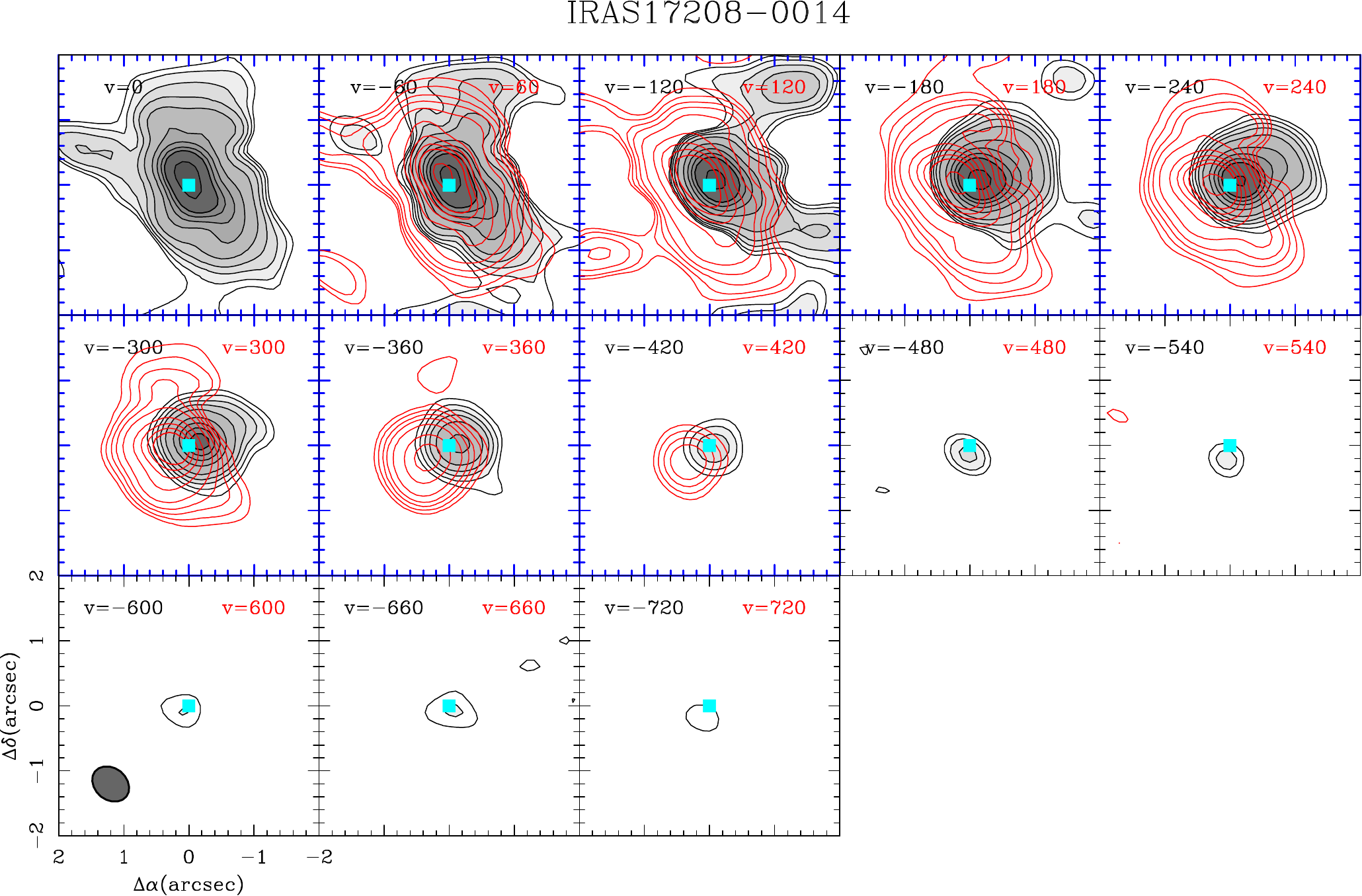}
    \caption{CO(2--1) velocity-channel maps observed with the PdBI in the nucleus of \object{IRAS~17208-0014} with a spatial resolution of  
$0\farcs6\times0\farcs5$ at  $PA=61^{\circ}$ (beam is plotted as a filled ellipse in the bottom left corner of the lower left panel). Velocity resolution is 60~km~s$^{-1}$. We show a field of view of 4$\arcsec$, i.e., $\sim$1/5 the diameter of the primary beam at 221.1~GHz.  We display channel maps grouped by pairs from 
$v-v_{\rm sys}$=0~km~s$^{-1}$ in steps of 60~km~s$^{-1}$ with  $v_{\rm sys}$(HEL)$~=12808$~km~s$^{-1}$ from --720~km~s$^{-1}$ to 720~km~s$^{-1}$. Emission at blueshifted velocities is displayed in gray scale and black contours, and emission at redshifted velocities is displayed in red contours. Contour levels are 3$\sigma$, 5$\sigma,$ and 8$\sigma$ with 1$\sigma=0.7$~mJy~beam$^{-1}$ for the channels of the line wing ($\mid$$v-v_{\rm sys}$$\mid$$~\geq 450$~km~s$^{-1}$), and  3$\sigma$, 5$\sigma$, 8$\sigma$, 12$\sigma$, 20$\sigma,$ to 100$\sigma$ in steps of  20$\sigma$, and 160$\sigma$ with 1$\sigma=1.4$~mJy~beam$^{-1}$ for the channels of the line core ($\mid$$v-v_{\rm sys}$$\mid$$~< 450$~km~s$^{-1}$; boxes highlighted in blue). The position of the dynamical center ([$\Delta\alpha$, $\Delta\delta$]~=~[0$\arcsec$,0$\arcsec$]~=~[$\alpha_{2000}=17^{h}23^{m}21.96^{s}$, $\delta_{2000}~=-00^{\circ}17^{\prime}00.87\arcsec$]) is highlighted by the (blue) square marker.}
                \label{IRAS17208-channels}
\end{figure*}
   
   
 \begin{figure*}[tbh!]
   \centering  
  \includegraphics[width=17cm]{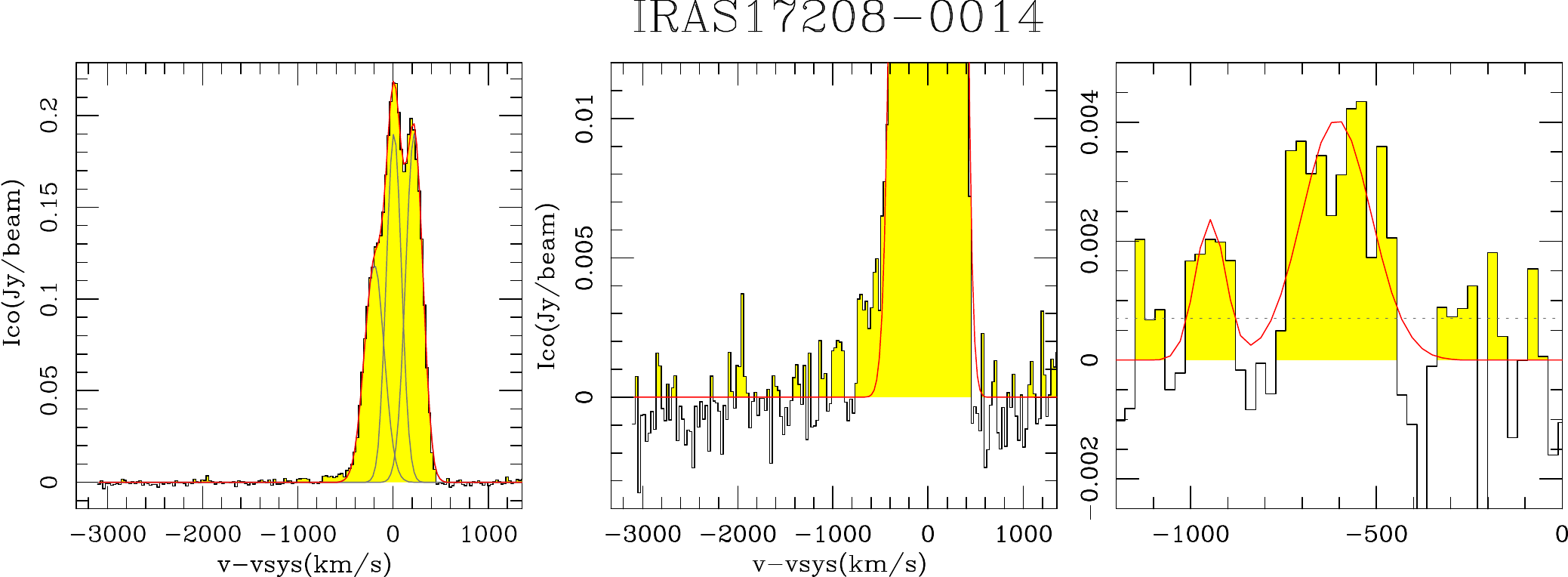}  
  
       \caption{Same as Fig.~\ref{spectra-1614} but for the CO(2--1) spectrum obtained toward the position that maximizes the emission in the line wing  in \object{IRAS~17208-0014} at $\alpha_{2000}=17^{h}23^{m}21.963^{s}$, $\delta_{2000}~=-00^{\circ}17^{\prime}01.03\arcsec$.}
              \label{spectra-17208}
\end{figure*}
   

 \begin{figure*}[tbh!]
   \centering  
  \includegraphics[width=17cm]{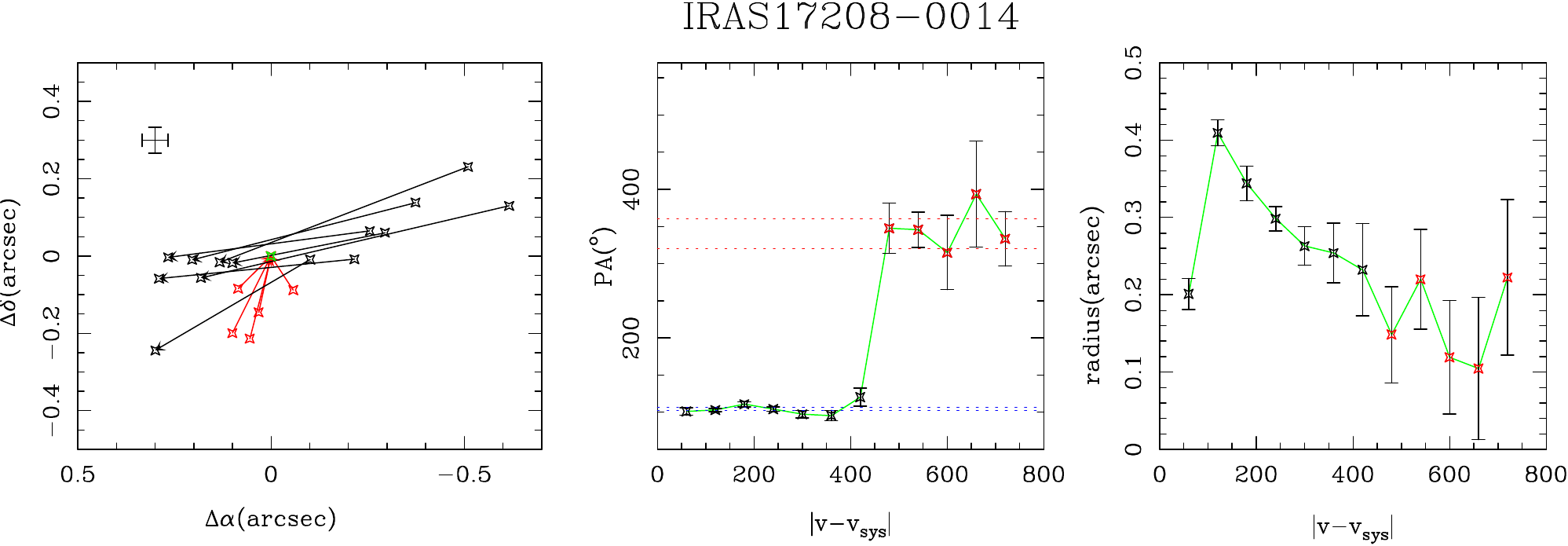}  
  
       \caption{Same as Fig.~\ref{PA-dist-1614} but for the centroids of CO(2--1) emission in \object{IRAS~17208-0014}.}
              \label{PA-dist-17208}
\end{figure*}
   

The middle panel of Fig.~\ref{1614-slices} shows the p-v plot taken along the kinematic minor axis ($PA=82^{\circ}$). The bulk of the emission from the line wings, which stems from the nuclear star-forming ring region, shows a distinct kinematic pattern: the high-velocity blueshifted (redshifted) gas is detected on the eastern (western) side of the ring up to $r\sim2\arcsec$~(0.6~kpc), as better shown in the close-up view of the right-hand panel of Fig.~\ref{1614-slices}. Taken at face value, the observed pattern invalidates the interpretation of the line wing emission as due to a fast-rotating coplanar disk.  A similar pattern can be
traced continuously into the lower velocities of the line core, which suggests that a single kinematic component is responsible for the velocity gradient
observed along the minor axis. For the assumed geometry of the disk, this implies that the gas would be moving {\it \emph{radially inward}} if it lay in the galaxy plane.
In particular, emission from the line wings would be coming from gas moving inward 
at very high velocities of up to $\sim(200-400)$/$sin(i)$~km~s$^{-1}$~=~370--740~km~$s^{-1}$. However, this  regime is well beyond the expected range
of inward velocities of a typical density-wave gas flow. 

In contrast, if we abandon the restriction of the coplanar solution for the gas, the kinematic pattern of the line wings can be satisfactorily explained either by a non-coplanar (less extreme) inflow solution or, alternatively, by a molecular outflow, depending on the adopted geometry.  As argued below, the non-coplanar outflow solution is favored. In the outflow scenario, the blueshifted velocities would stem from the lobe pointing toward us on the eastern side of the disk, while the redshifted velocities would be behind the disk on the western side. This outflow model for CO agrees perfectly with the geometry of the H$\alpha$ outflow discovered by Bellocchi et al.~(\cite{Bel12}). The non-detection of redshifted H$\alpha$ emission indicates that this component is located behind the disk, and therefore remained undetected owing to extinction. The extinction derived from the 1.6$\mu$m/2.2$\mu$m color map of Alonso-Herrero et al.~(\cite{Alo01}) appears to be higher on the western side of the ring.  The similar spatial extent and (blueshifted) velocities measured for the H$\alpha$ and CO features suggest that both are part of the same outflowing component. The observed tilt in the major axis of the molecular outflow relative to the $PA$ of the rotating disk ($\sim80^{\circ}$ as discussed in Sect.~\ref{1614-ch}) suggests that the outflow has a significant component  that is oriented perpendicular to the rotating disk. This geometry is typically found in star-forming-driven molecular outflows, like the ones discovered in the starbursts M~82 and NGC~253 (Garc\'{\i}a-Burillo et al.~\cite{Gar01}; Walter et al.~\cite{Wal02}; Bolatto et al.~\cite{Bol13b}; Salas et al.~\cite{Sal14}). However, we cannot exclude that an AGN can also contribute to drive the outflow (see discussion in Sect.~\ref{out-power}).

\subsection{\object{IRAS~17208-0014}: channel maps and emission centroids}\label{17208-ch}


Figure~\ref{IRAS17208-channels} shows the CO(2--1) velocity-channel maps observed in \object{IRAS~17208-0014} with a spatial
resolution of  $0\farcs6\times0\farcs5$ at  $PA=61^{\circ}$, obtained using uniform weighting of the data.  We display channel maps grouped by pairs with a velocity spacing of 60~km~s$^{-1}$ from $v-v_{\rm sys}=0$~km~s$^{-1}$ with  
$v_{\rm sys}$(HEL)$~=12808$~km~s$^{-1}$, a value determined by the best-fit solution described in Sect.~\ref{17208-mod}.

Most of the CO  emission stems from the central $r \sim 2\arcsec$ (1.7~kpc) region shown in Fig.~\ref{IRAS17208-channels}.  The emission covers a wide and remarkably  asymmetric velocity range: $v-v_{\rm sys}$~=~[+450,--750]~km~s$^{-1}$. However, up to $\simeq 99\%$ of the total emission comes from a narrower and symmetric velocity range: $v-v_{\rm sys} \sim [+450,-450]$~km~s$^{-1}$. Following the notation adopted in \object{NGC~1614}, we also refer to this central velocity interval as the line core and identify the emission outside this range as coming from the line wing (i.e., $-450$~km~s$^{-1}$$> v-v_{\rm sys} \geq -750$~km~s$^{-1}$). Although at an admittedly lower level compared to the line core, emission from the line wing reaches an overall statistical significance of about $\sim15\sigma$ in velocity-integrated units.

Figure~\ref{spectra-17208} shows the CO emission profile extracted from the position where the emission in the line wing is maximum at [$\Delta\alpha$, $\Delta\delta$]~=~[0$\arcsec$,--$0\farcs2$], which corresponds to $\alpha_{2000}=17^{h}23^{m}21.963^{s}$, $\delta_{2000}~=-00^{\circ}17^{\prime}01.03\arcsec$.
We used three Gaussian components  to fit the emission from the line core. As in \object{NGC~1614}, the fit left  a residual of emission, which  in  \object{IRAS~17208-0014} is identified at highly blueshifted velocities in the range $v-v_{\rm sys}$~=~[--450,--750]~km~s$^{-1}$ and a tentatively detected component around --950~km~s$^{-1}$. This blueshifted  line wing has no redshifted counterpart, unlike in 
\object{NGC~1614}.  Emission from the line core reveals a rotating disk with a kinematic 
major axis oriented at a small angle relative to  the east-west axis:  emission at redshifted (blueshifted) velocities comes from the eastern (western) side of the disk, as illustrated in Fig.~\ref{IRAS17208-channels}. Within the limits of our spatial resolution, Fig.~\ref{IRAS17208-channels} shows that emission in the line wing shifts progressively from the western to the eastern side of the disk as the velocity offset increases, i.e., showing an opposite trend compared to the line core. This is a signature of kinematic decoupling of the gas emitting at high velocities.

We have derived the centroids  of CO emission in \object{IRAS~17208-0014} to quantify  the structure of the different kinematic components following the same
procedure as described in Sect.~\ref{1614-ch} for \object{NGC\,1614}. Figure~\ref{PA-dist-17208} shows both the centroids and the $PA$ and $r$ profiles obtained for the same velocity channels as are displayed in Fig.~\ref{IRAS17208-channels}. As shown in the middle panel of Fig.~\ref{PA-dist-17208}, the value of $PA$ stays fairly constant inside the line core around $PA_{\rm core}=105\pm2^{\circ}$, an angle that is close to the value derived in Sect.~\ref{17208-mom} for the $PA$ of the rotating disk ($PA\sim 113^\circ\pm 3^\circ$). In stark contrast, the $PA$ values for the line wing centroids oscillate around a weighted mean of $PA_{\rm wing}=340^{\circ}\pm20^{\circ}$, an indication that its apparent  kinematic major axis is virtually reversed: the $PA$ of the line wing changes by an angle $\geq230^{\circ}$ relative to the line core \footnote{In the absence of a redshifted counterpart in the line wing, we derived the $PA$ for this component  using the central offset and the blueshifted channel centroids as reference.}.  We conclude that the evidence of
a kinematic decoupling, reflected in the measured difference $PA_{\rm wing}-PA_{\rm core}\sim235\pm20^{\circ}$ is a significant $\simeq10\sigma$ result. 
Furthermore, the line wing channels show lower $r$ values ($\sim0\farcs1-0\farcs2$~($80-160$~pc)) compared to the   line core (up to $\sim0\farcs3$~(240~pc)), indicating that the gas emitting at anomalous velocities lies at comparatively smaller radii.

 \begin{figure*}[tbh!]
   \centering  
  \includegraphics[width=17cm]{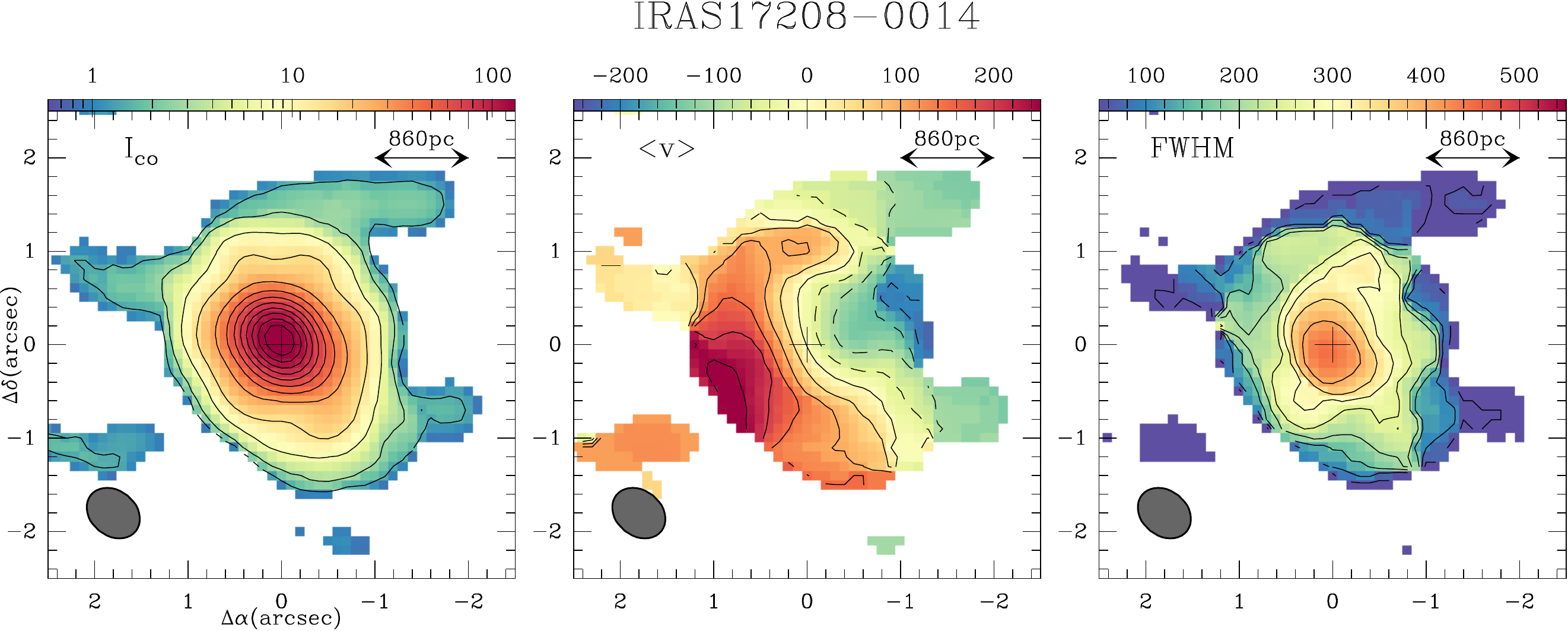}
   \caption{{\it Left panel}:~The CO(2--1) integrated intensity map obtained in the disk of \object{IRAS~17208-0014}. The map is shown in color scale with contour levels  1$\%$, 2$\%$, 5$\%$, 10$\%,$ to  90$\%$ in steps of 10$\%$ of the peak value~=~143~Jy~km~s$^{-1}$beam$^{-1}$ ($\sim$300~$\sigma$). {\it Middle panel}:~The CO(2--1) isovelocity contours spanning the range (--250~km~s$^{-1}$, 250~km~s$^{-1}$) in steps of 50~km~s$^{-1}$ are overlaid on a false-color velocity map (linear color scale as shown). Velocities refer to $v_{\rm sys}$(HEL)$~=12808$~km~s$^{-1}$. {\it Right panel}:~Overlay of the CO(2--1) line widths (FWHM) shown in contours (50 to 550~km~s$^{-1}$ in steps of 50~km~s$^{-1}$) on a false-color width map (linear scale as shown). The filled
ellipses in the bottom  left corner of each panel represent the CO(2-1) beam size ($0\farcs6\times0\farcs5$ at  $PA=61^{\circ}$). The position of the dynamical center  is highlighted by the cross marker.}         
 \label{IRAS17208-moments}
\end{figure*}
   

 \begin{figure*}[tbh!]
   \centering  
  \includegraphics[width=17cm]{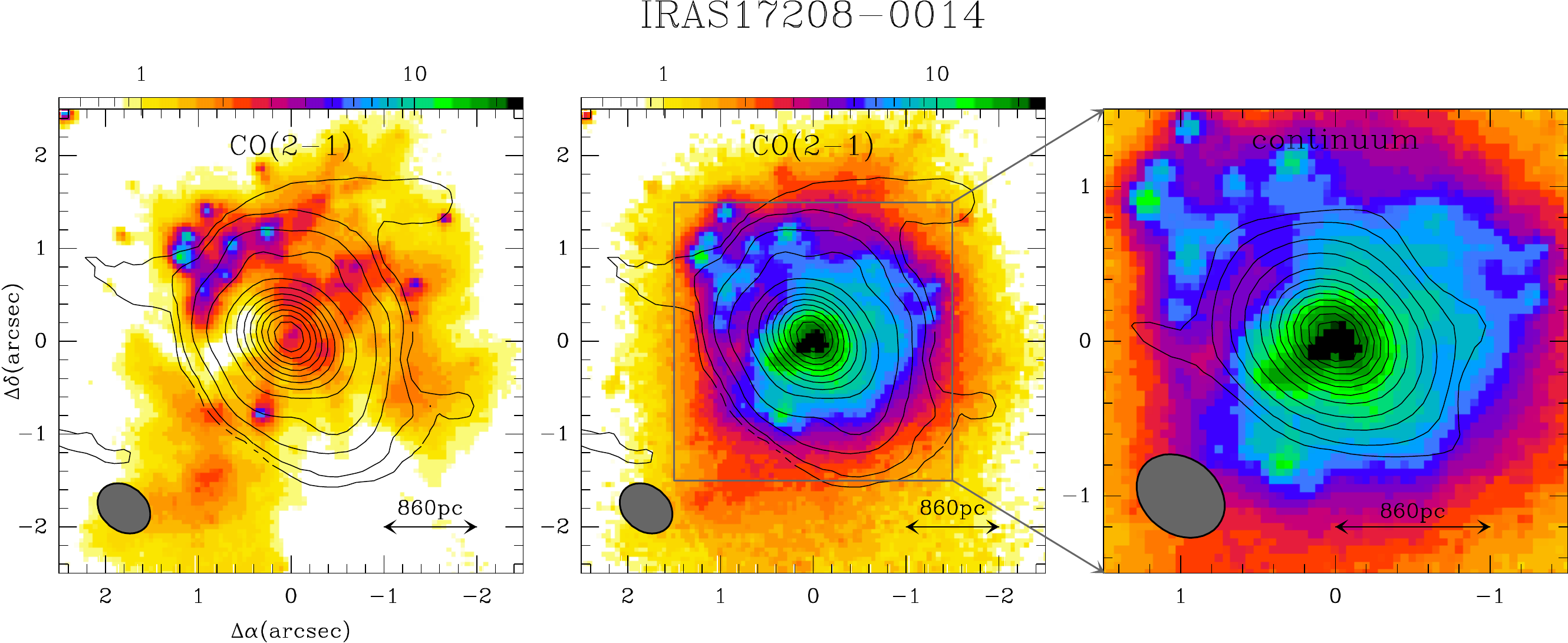}
   \caption{{\it Left panel}:~Overlay of the CO(2--1) intensity contours (levels as in the left-hand panel of Fig.~\ref{IRAS17208-moments}) on the  HST optical 
   image of \object{IRAS~17208-0014} obtained in the F814W band (color scale as shown in arbitrary units). {\it Middle panel}:~Same as {\it left 
   panel} but with CO contours overlaid on the HST image obtained with NICMOS in the F160W band  (color scale as shown in arbitrary units). 
   {\it Right panel}:~Same as {\it middle panel} but zooming in on the continuum emission image of \object{IRAS~17208-0014} at 221.1~GHz, in 
   contours (1$\%$, 2$\%$, 5$\%$, 10$\%,$ to  90$\%$ in steps of 10$\%$ of the peak value~=~30.4~mJy~beam$^{-1}\sim250\sigma$), overlaid on 
   the HST NICMOS image. The filled ellipses represent the CO beam size.}
              \label{IRAS17208-overlay}
\end{figure*}
   

\subsection{\object{IRAS~17208-0014}: moment maps}\label{17208-mom}

Figure~\ref{IRAS17208-moments} shows the moment maps derived from the CO(2--1) line data  inside the velocity interval  $v-v_{\rm sys}$~=~[-1000, 500]~km~s$^{-1}$. 
As for \object{NGC~1614}, we used different clippings: a $3\sigma$-clipping on the intensities to derive the zero-order moment (left-hand panel of Fig.~\ref{IRAS17208-moments})
and a $5\sigma$-clipping to derive the mean-velocity field and the velocity width maps (shown in  the middle and right-hand panels of Fig.~\ref{IRAS17208-moments}, respectively).

The left-hand panel of Fig.~\ref{IRAS17208-moments} shows that the CO emission comes from a spatially-resolved  $3\farcs3 \times 2\farcs2$ (2.7~kpc~$\times$~1.8~kpc) molecular disk feature that is noticeably elongated along $PA\sim20^{\circ}$.  
The overall morphology of the CO disk is similar to the northeast-southwest elongated structure seen in the dust extinction, as derived from the 2.2$\mu$m/1.1$\mu$m HST color image of Scoville et al.~(\cite{Sco00}).  However, compared to the CO(3--2) disk of Wilson et al.~(\cite{Wil08}), the CO(2--1) disk is a factor 1.5--2 larger.  As shown in the mean-velocity field of the middle panel of Fig.~\ref{IRAS17208-moments}, which confirms the signature of a spatially resolved rotating disk, the morphological and kinematical major axes of the CO disk are roughly orthogonal. This implies that the intrinsic elongation of the disk should be more pronounced after deprojection.  Three fainter protrusions and a detached clump extend the disk emission farther out up to $r\sim2.2\arcsec$~(1.8~kpc). Most of these CO extensions coincide in position with similar features of the 2.2$\mu$m/1.1$\mu$m HST color image of Scoville et al.~(\cite{Sco00}).

If, as discussed by Arribas \& Colina~(\cite{Arr03}), the spiral
arms and the most prominent tidal feature of the optical images of the galaxy (e.g., Melnick \& Mirabel~\cite{Mel90}; Solomon et al.~\cite{Sol97}) are assumed to be trailing, gas rotation should be counterclockwise, and the observed CO velocity field would imply that the northeastern side of the disk is the near side. The ($0\arcsec,0\arcsec$)--offset, initially adopted as the nucleus of the galaxy,  maximizes the symmetry of the mean-velocity field. It also corresponds to the maximum velocity width (FWHM) measured in the disk, as shown in the right-hand panel of Fig.~\ref{IRAS17208-moments}, and, also, to the peak of the continuum emission at 221.1~GHz, as shown in 
the right-hand panel of Fig.~\ref{IRAS17208-overlay}. Therefore we adopt  this position as the dynamical center of the galaxy.

 The average FWHM  of the CO line in the disk is $\sim$170~km~s$^{-1}$, and this is lowered to $\sim$150~km~s$^{-1}$ if we exclude the central $r=0\farcs3$~(0.2~kpc) of the disk to 
 minimize the effects of beam smearing. This would imply an anomalously high average velocity dispersion of $\sigma \simeq 64$~km~s$^{-1}$. However, the same estimate
 is a factor of two lower if we derive it in the outer disk  region ($r>1\farcs5$~(1.2~kpc)), an indication that there are unresolved gradients of circular and non-circular motions in the
 disk. According to the middle panel of Fig.~\ref{IRAS17208-moments},  the kinematic major axis of the rotating disk is oriented along  $PA\sim 110^{\circ}$. 
However, there are significant departures from circular rotation, identified by wiggles in the velocity field  at several locations in the disk.

 \begin{figure*}[tbh!]
   \centering  
  \includegraphics[width=17cm, angle=0]{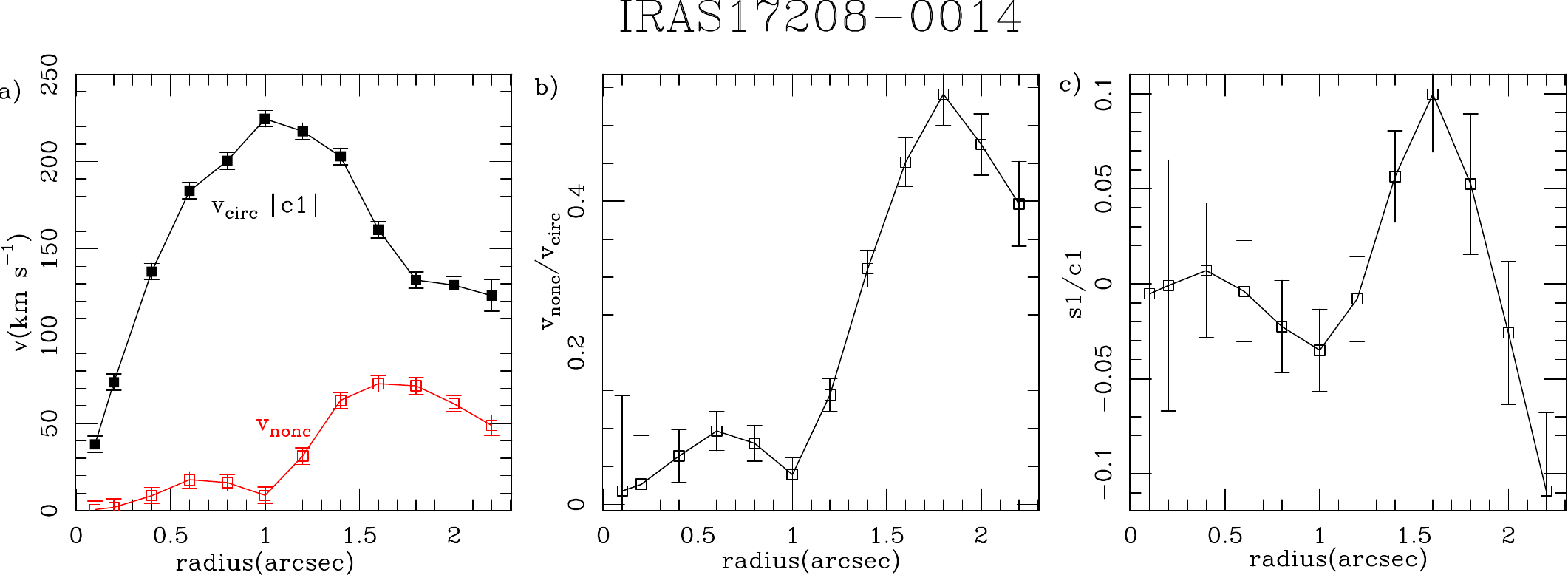}
  
       \caption{Same as Fig.~\ref{1614-kinemetry}, but derived from the Fourier decomposition of the velocity field of \object{IRAS~17208-0014}.}
              \label{17208-kinemetry}
\end{figure*}
   

 \begin{figure*}[tbh!]
   \centering  
  \includegraphics[width=11.5cm, angle=0]{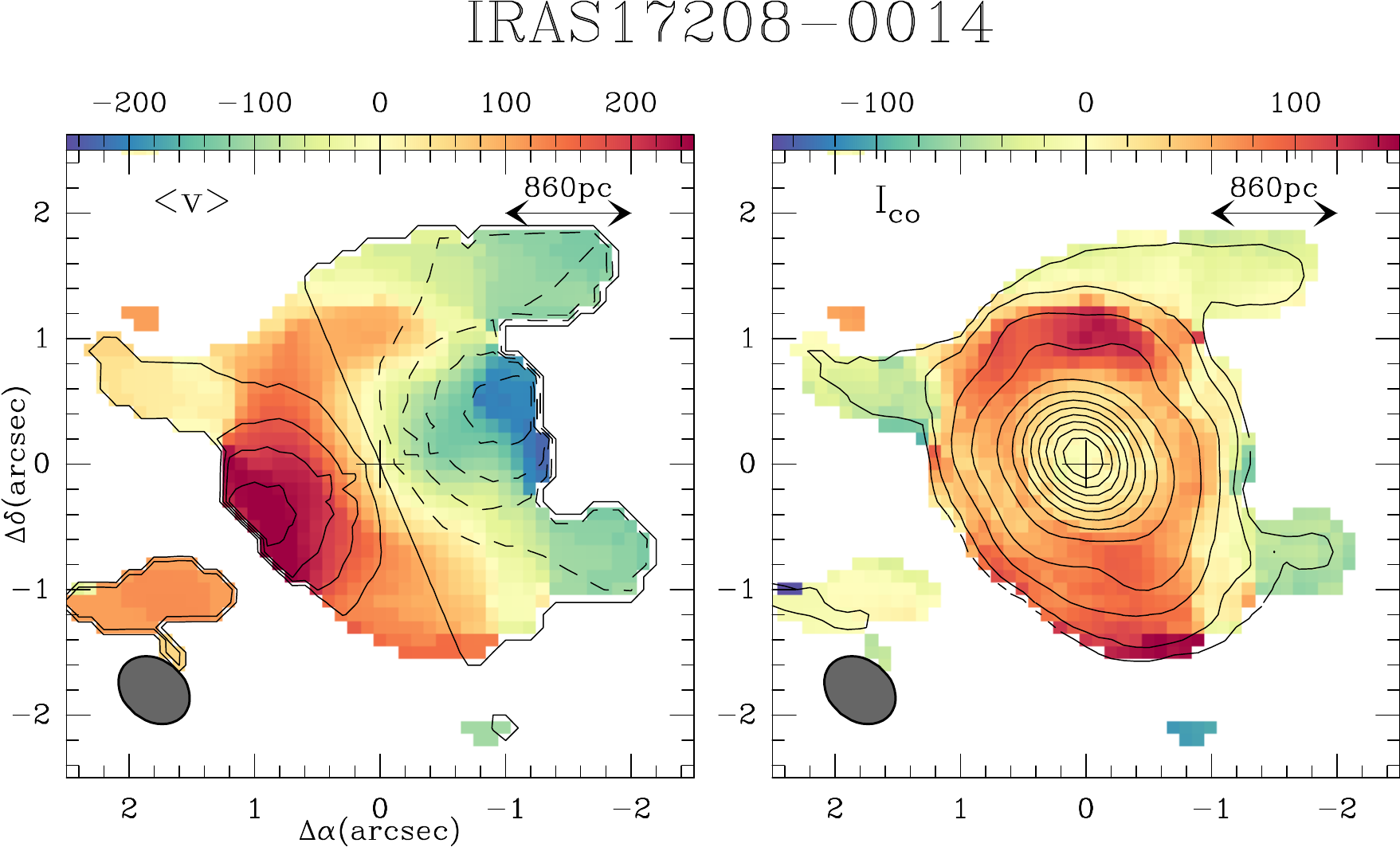}
     
       \caption{Same as Fig.~\ref{1614-kinemetry-residuals}, but showing here the comparison between the CO(2--1) observations and the best-fit model in \object{IRAS~17208-0014}.
Contours in the {\it left panel} span the range --200~km~s$^{-1}$ to 200~km~s$^{-1}$ in steps of 50~km~s$^{-1}$. The CO(2--1) intensity contours in the {\it right panel} are as in the left-hand panel of Fig.~\ref{IRAS17208-moments}. Color scales in both panels are as shown.}
              \label{17208-kinemetry-residuals}
\end{figure*}
   

The left-hand and middle panels of Fig.~\ref{IRAS17208-overlay} overlay the CO line intensities on the HST  optical and NIR images of \object{IRAS~17208-0014} obtained in the F814W and F160W 
bands and published, respectively, by Arribas \& Colina~(\cite{Arr03}) (from program 6346 by PI: K. Borne) and Scoville et al.~(\cite{Sco00}). The bulk of the CO emission is associated  with the compact starburst region, 
which hosts $\simeq18$ massive star clusters cataloged by Scoville et al.~(\cite{Sco00}).  Unlike the CO line emission, the mm-continuum emission at 221.1~GHz, yet spatially resolved\footnote{The 
continuum emission has a deconvolved size of $0\farcs3 \times 0\farcs2$, determined by task UVFIT of GILDAS.}, is mostly restricted to the compact starburst region, as 
shown in the right-hand panel of Fig.~\ref{IRAS17208-overlay}.

\subsection{\object{IRAS~17208-0014}: kinematic modeling}\label{17208-mod}

\subsubsection{{\tt Kinemetry fit}: the coplanar solution}\label{17208-kin}
     
As for \object{NGC~1614}, we used the CO mean-velocity field of \object{IRAS~17208-0014} to find the best-fit Fourier decomposition of the line-of-sight velocities using {\tt kinemetry}   (Krajnovi{\'c} et al.~\cite{Kra06}). Figure~\ref{17208-kinemetry} shows the radial profiles of the main terms of the Fourier decomposition, and Fig.~\ref{17208-kinemetry-residuals} shows the rotating disk solution of {\tt kinemetry} and the residuals of the fit. As expected, circular rotation accounted for by the $c_1$ term is dominant throughout the disk
(Fig.~\ref{17208-kinemetry}a). The best-fit geometry for the rotating disk is defined by a position angle $PA=113\pm3^{\circ}$  and an inclination $i=55\pm3^{\circ}$. We also determined $v_{\rm sys}$(HEL)$~=12808\pm15$~km~s$^{-1}$ as part of the fit. 
Relative to $v_{\rm circ}$, non-circular motions ($v_{\rm nonc}$) are within a 2\%\ to 50$\%$ range with a clear trend that shows higher non-circular motions at larger radii (Fig.~\ref{17208-kinemetry}b). Furthermore, based on the sign of the $s_{1}$ term, the pattern of the purely radial motions is undefined in the central $r \sim 1\farcs2$~(1~kpc) region, where the $s_{1}/c_{1}$ ratio shows an oscillating profile compatible with zero. Outside this central region, motions are indicative of a moderate outflow in the radial range $r \sim 1\farcs2-1\farcs9$~(1--1.5~kpc), while they suggest inflow farther out at $r > 1\farcs9$~(1.5~kpc). Deprojected, these motions would translate into a maximum amplitude of $\leq25$~km~s$^{-1}$. The $s_1$ term in \object{IRAS~17208-0014} is nevertheless a small fraction ($\leq0.2-0.3$) of $v_{\rm nonc}$ in the outer molecular disk ($r>1\farcs5$~(0.8~kpc)).

\subsubsection{A non-coplanar outflow solution}\label{17208-outflow}  

Overall, similar to \object{NGC~1614}, the order of magnitude of the fitted noncircular motions can be explained as due to density waves. However, this scenario, which implicitly assumes that gas is in a coplanar geometry, is unable to explain the order of magnitude or even the {\em \emph{asymmetric}} pattern of the  high velocities ascribed to the line wing, as argued below.

\begin{figure*}[tbh!]
   \centering  
  \includegraphics[width=17cm]{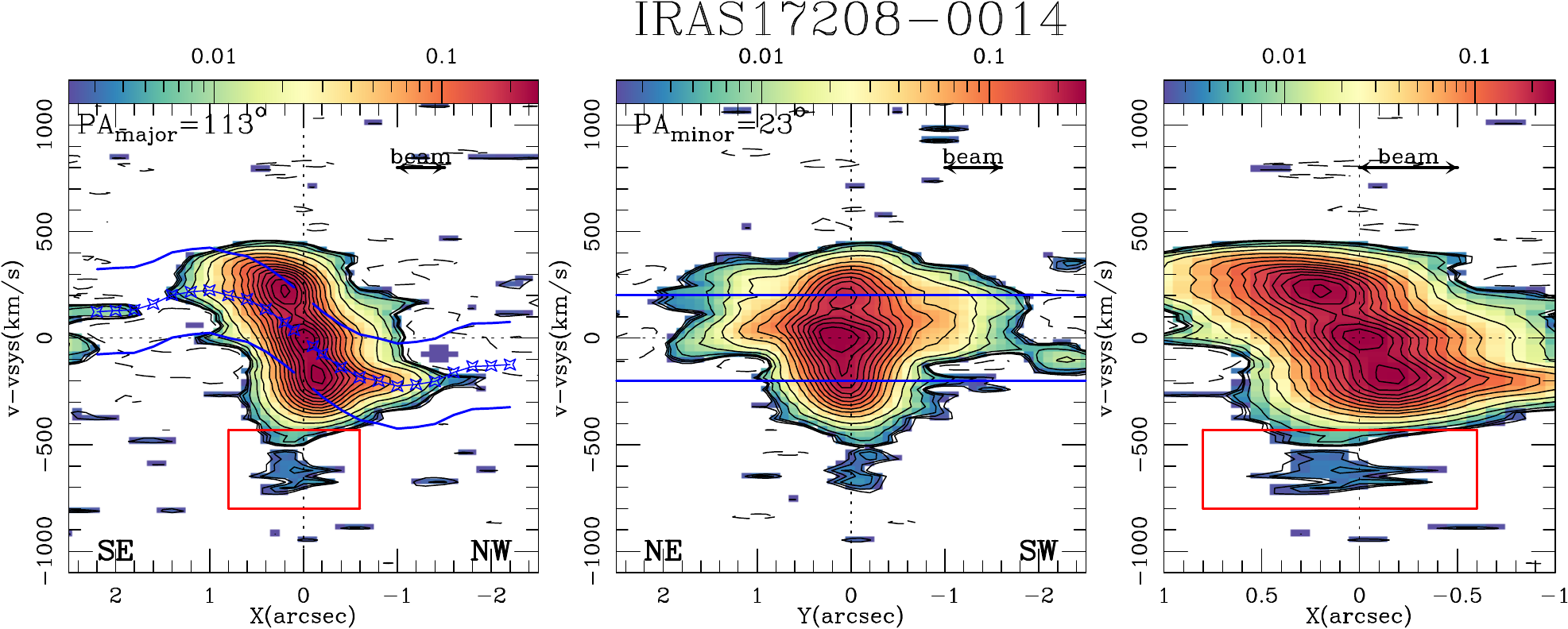}  
       \caption{{\it Left panel}:~The (p-v) plot taken along the kinematic major axis ($PA=113^{\circ}$) of \object{IRAS~17208-0014}. As in Fig.~\ref{1614-slices}, we delineate  the (projected) rotation curve and the allowed virial range around it.  Similarly, we delimit the region where emission is detected from the line wing.
 {\it Middle panel:}~The p-v plot taken along the kinematic minor axis ($PA=23^{\circ}$). As in Fig.~\ref{1614-slices}, the allowed virial range is delimited by the blue curves. {\it Right panel:}~Same as  {\it left panel} but zooming in on the central $\pm1\arcsec$ region around the center.      
In all panels, contour levels are -2$\sigma$, 2$\sigma$, 2.5$\sigma$, 3$\sigma$, 5$\sigma$, 10$\sigma$, 20$\sigma,$ to 220$\sigma$ in steps of 20$\sigma$ with 1$\sigma=1.2$~mJy~beam$^{-1}$. The velocity scale is relative to $v_{\rm sys}$. The spatial 
scales (X and Y) are relative to the dynamical center. We highlight the spatial resolution (beam size) achieved with UN weighting in each panel.}
              \label{17208-slices}
\end{figure*}

The left-hand panel of Fig.~\ref{17208-slices} shows the major axis p-v plot (along $PA=113^{\circ}$) with the projected best-fit rotation curve superposed. 
A sizeable fraction of the emission in the line core lies within the {\it \emph{virial}} range determined by a combination of circular rotation ($\sim \pm230$~km~s$^{-1}$), turbulence (FWHM/2$\sim \pm85$~km~s$^{-1}$), and an upper limit to the likely  contribution from in-plane non-circular motions\footnote{This amounts to $\pm 115$~km~s$^{-1}$ in \object{IRAS~17208-0014}.}.  While the left-hand panel of Fig.~\ref{17208-slices} shows that the model, limited by beam smearing effects, may have underestimated the rotation curve gradient in the central $r\leq0\farcs5$~(0.4~kpc) region, a hypothetically steeper rotation curve would nevertheless fail to explain the line wing feature: the stark asymmetry of the line wing, only detected at blue velocities, argues against rotational support as an explanation of high velocities. Furthermore, as illustrated best in the right-hand panel of Fig.~\ref{17208-slices},  with the achieved spatial resolution we locate the line wing feature in a quadrant of the p-v plot that is formally assigned to gas in (coplanar) counter-rotation. Nevertheless, the coexistence of two opposite spin gas components at the same radii is highly unlikely because of the intrinsic dissipative nature of  the gas.

The alternative non-coplanar solution for the line wing in \object{IRAS~17208-0014} can be used to invoke either  inflow or outflow, depending on the adopted geometry.  However, the detection by Arribas et al~(\cite{Arr14}) of a similar, yet less extreme, blueshifted  H$\alpha$ component in \object{IRAS~17208-0014} favors  the outflow scenario for the  line wing, similar to the case of \object{NGC~1614} discussed in Sect.~\ref{1614-mod}. In addition, the atomic and molecular outflows seen, respectively, in the Na{\small I}~D line (Rupke \& Veilleux~\cite{Rup13}) and in the OH line (Sturm et al.~\cite{Stu11}) give further support to the outflow scenario as an explanation for the CO extreme velocities in  \object{IRAS~17208-0014}.

\section{Basic properties of the molecular outflows of  \object{NGC~1614} and \object{IRAS~17208-0014}}\label{out-properties}

\subsection{Morphology}\label{out-mor}

\begin{figure*}[tbh!]
   \centering  
  \includegraphics[width=7cm]{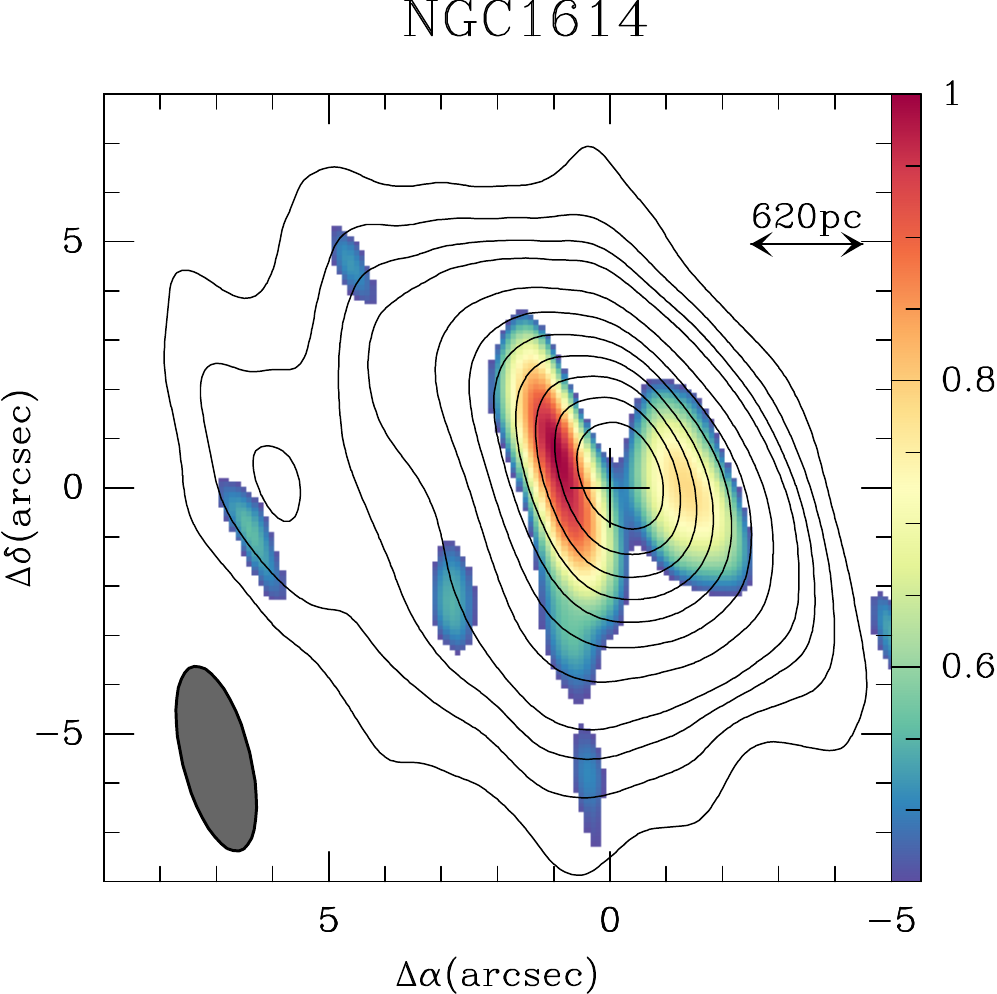}
  \includegraphics[width=6.9cm]{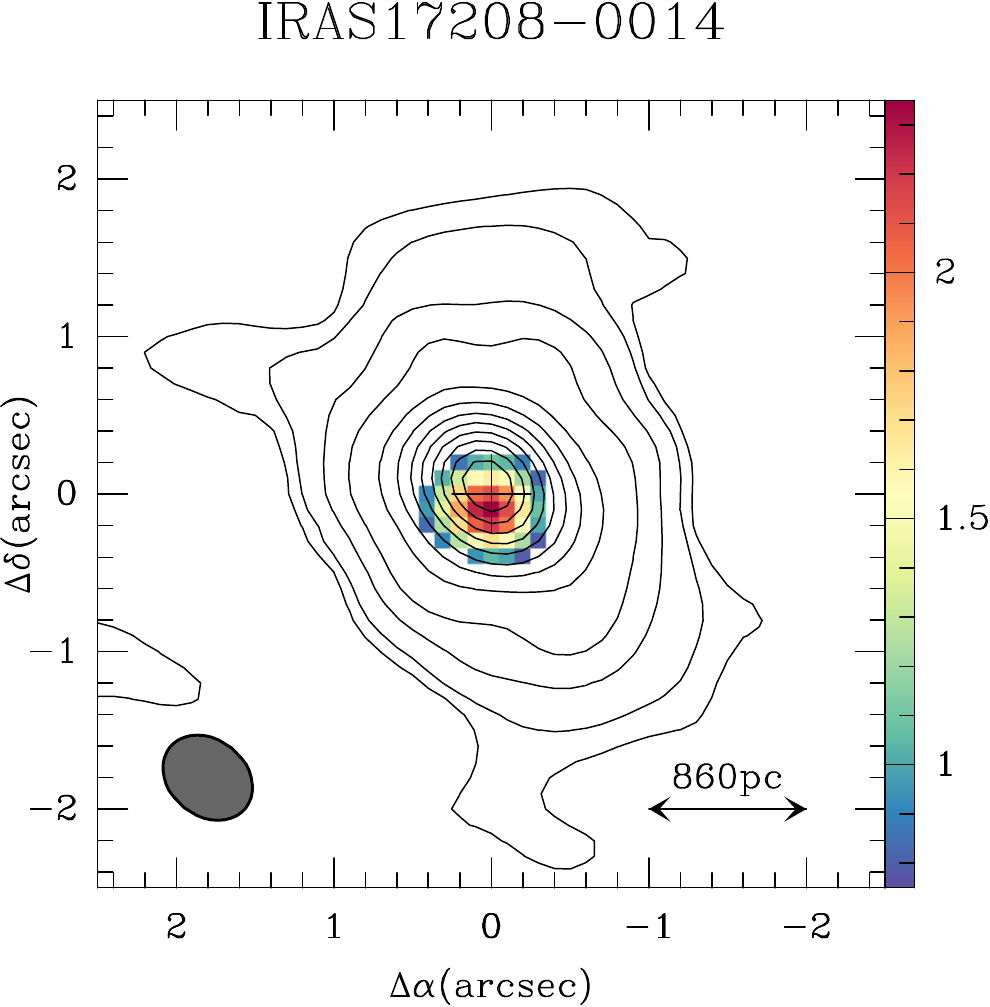}
  
       \caption{{\it Left panel:}~Overlay of the CO(1--0) intensity contours of the disk of \object{NGC~1614}, obtained by integrating the emission of the NA weighting data set inside the line  core, on the intensity map of the molecular outflow (in color), obtained by integrating the emission inside the line wing region. The two velocity intervals are defined in Sect.~\ref{1614-ch}.  Color scale spans the range [5$\sigma$,11$\sigma$], and contour levels are: $5\%$, $10\%$,  $15\%$, $20\%,$  to $90\%$  in steps of 
       10$\%$ of the peak value~=~38~Jy~km~s$^{-1}$~beam$^{-1}$.  {\it Right panel:}~Same as {\it left panel} but obtained from the UN weighting data set  of  \object{IRAS~17208-0014}. The color scale spans the range [5$\sigma$,15$\sigma$], and contour levels are $1\%$, $2\%$,  $5\%$, $10\%,$  to $90\%$  in steps of 
       10$\%$ of the peak value~=~127~Jy~km~s$^{-1}$~beam$^{-1}$. The two velocity intervals are defined in Sect.~\ref{17208-ch}.}
              \label{outflow-mom}
\end{figure*}

Figure~\ref{outflow-mom} shows the CO intensity maps of the molecular outflows in  \object{NGC~1614} and \object{IRAS~17208-0014}, derived by integrating the emission of the line wing components defined, respectively, in Sects.~\ref{1614-ch} and \ref{17208-ch}. We cannot rule out that emission at lower radial velocities apparently within the virial range is not coming also from outflowing gas seen at a larger angle relative to the line of sight. However, we purposely adopt here the most conservative definition of the outflows by restricting their velocity ranges to the most extreme values associated with the line wings (see however discussion in Sect.~\ref{17208-extreme}).

The molecular outflow in  \object{NGC~1614} consists mainly of two barely resolved  knots of emission located close to the nucleus along $PA\sim90^{\circ}$ at 
($\Delta\alpha$, $\Delta\delta$)=($+0\farcs8\sim250$~pc, $0\arcsec$) and  ($-1\farcs5\sim-470$~pc, $0\arcsec$) (see the left-hand panel of Fig.~\ref{outflow-mom}). The east (west) knot  corresponds to the blueshifted (redshifted) lobe of the outflow. The knots are bridged by lower level emission at the center. This morphology is reminiscent of a biconical outflow uniformly filled by the outflowing gas. Overall, the bulk of the molecular outflow is cospatial with the $2\arcsec$-diameter Pa$\alpha$ star-forming ring imaged by Alonso-Herrero et al.~(\cite{Alo01}). However,  the H$\alpha$ emission coming from the blueshifted lobe of the outflow, identified by Bellocchi et al.~(\cite{Bel12}), extends farther out up to $r\sim5\arcsec$~(1.6~kpc) on the eastern side of the disk. In this region, there is tentative evidence of outflowing CO emission that stems from a number ($\sim4$) of clumps. The velocity range of the outflow is comparable in CO and H$\alpha$:  the approaching outflow lobe is blueshifted by $\sim200-500$~km~s$^{-1}$ relative to $v_{\rm sys}$ in both tracers. 

  The molecular outflow in \object{IRAS~17208-0014}, identified as a blueshifted component, is associated with an off-centered knot of barely resolved emission located at ($\Delta\alpha$, $\Delta\delta$)=(0, $-0\farcs2\sim-160$~pc) (see the right-hand panel of Fig.~\ref{outflow-mom}). Within the errors this coincides with the position of the peak H$_2$ emission mapped by Medling et al.~(\cite{Med14}), attributed to shocked gas at the base of an outflow coming from the nucleus of one of the two stellar disks (Medling et al.~\cite{Med15}). In contrast, the neutral atomic outflow traced by the blueshifted Na{\small I}~D absorption comes from a wide angle ($\geq50^{\circ}$) and spatially extended component (up to $r\geq2\arcsec\sim1.6$~kpc in the plane of the sky) (Rupke \& Veilleux~\cite{Rup13}). Overall, the morphology of the atomic outflow is elongated along the CO kinematic major axis with hardly any extension along the minor axis, mimicking a wide angle (single) conical outflow. The CO outflow shows radial velocities similar to those measured for the atomic gas in the inner $r\sim1\arcsec$ (0.8~kpc): velocities are blueshifted by $\sim600$~km~s$^{-1}$ relative to $v_{\rm sys}$ in both tracers. However, the (blueshifted) velocities of the H$_2$ outflow measured by  Medling et al.~(\cite{Med15}) are $\sim$a factor of 5 lower than the CO values. In Sect.~\ref{17208-extreme} we propose a redefinition of the line wing velocities, which results in a change in the morphology, mass, and energetics of the CO outflow.


\begin{table*}[hbt!]
\caption{\label{t1}Molecular outflow properties.}
\centering
\begin{tabular}{lccccccccccc}
\noalign{\smallskip} 
\hline
\hline
\noalign{\smallskip} 
Sources & $R_{\rm out}$ & $V_{\rm out}$ & $M_{\rm out}$ & $\dot{M}_{\rm out}$ & $ \frac{\dot{M}_{\rm out}}{SFR}$      & $L_{\rm out}$ & $\frac{L_{\rm out}}{L_{\rm SNe}}$ &  $\frac{L_{\rm out}}{L_{\rm AGN}}$ & $\dot{P}_{\rm out}$ &     $\frac{\dot{P}_{\rm out}}{L_{\rm bol}/c}$ & $\frac{\dot{P}_{\rm out}}{L_{\rm AGN}/c}$        \\
\noalign{\smallskip}
              &  (pc) & (km~s$^{-1}$)   &       ($M_{\sun}$)        &   ($M_{\sun}$~yr$^{-1}$)  &       --      &       ($L_{\sun}$)    & --      &    -- &       (g~cm~s$^{-2}$) & -- & -- \\
\noalign{\smallskip}                  
\hline
\noalign{\smallskip}          
 \object{NGC~1614}        & 560  &      360     & $3.2\times10^{7}$        &     40      &       0.8     &       $6.5\times10^{8}$ & 0.06        &       $>0.05~$\tablefootmark{a}       & $1.4\times10^{35}$   &  1.9    & $>65~$\tablefootmark{a}  \\
 \noalign{\smallskip}  
\hline 
\noalign{\smallskip} 
 \object{IRAS~17208}  & 160  &  600     & $4.6\times10^{7}$        &    330     &       $1.4~$\tablefootmark{b} &       $1.6\times10^{10}$ & $0.4~$\tablefootmark{b}    &       $>6~$\tablefootmark{c}  & $1.6\times10^{36}$   &  5    & $>5\times10^{3}$\tablefootmark{c}       \\
        &  -- & --      &  --   &  --   & $2.0~$\tablefootmark{d}  & -- & $0.6~$\tablefootmark{d}  &      $0.02~$\tablefootmark{e}        &  --  &   --   & $17~$\tablefootmark{e}         \\
    
 \noalign{\smallskip}    
\hline
\hline 
\end{tabular} 
\tablefoot{All properties are derived for the line wing components as defined in Sects.~\ref{1614-ch} and \ref{17208-ch} assuming $\alpha=33^{\circ}$ and 35$^{\circ}$ in  Eqs.~(\ref{out-1}), (\ref{kin}), and (\ref{mom}) for \object{NGC~1614} and \object{IRAS~17208-0014}, respectively.  
Outflow mass rates, momentum rates, and  kinetic luminosities have uncertainties of $\pm0.34$~dex, $\pm0.36$~dex, and  $\pm0.40$~dex, respectively. The mass, momentum, and energy loading factors have associated uncertainties of $\pm0.43$~dex. See Sect.~\ref{out-power} for details.
\\
\tablefoottext{a}{Derived assuming $L_{\rm AGN}^{\rm NGC~1614}<1.3\times10^{10}~$L$_{\sun}$     (Pereira-Santaella et al.~in prep.).} \\
\tablefoottext{b}{Derived assuming $SFR=240~M_{\sun}$~yr$^{-1}$.} \\
\tablefoottext{c}{Derived assuming $L_{\rm AGN}^{\rm IRAS~17208}<2.4\times10^{9}~$L$_{\sun}$     (Gonz\'alez-Mart\'{\i}n et al.~\cite{Gon09}).}\\
\tablefoottext{d}{Derived assuming $SFR=0.7\times240=168~M_{\sun}$~yr$^{-1}$.}  \\
\tablefoottext{e}{Derived assuming $L_{\rm AGN}^{\rm IRAS~17208}\sim7.2\times10^{11}~$L$_{\sun}$    (Aalto et al.~\cite{Aal15b}.).}}\\
\end{table*}

\subsection{Mass, energy, and momentum rates} \label{out-load}

To quantify the mass load budget ($\dot{M}_{\rm out}\equiv$d$M_{\rm out}$/d$t$), we use the expression 
 
\begin{equation}
\dot{M}_{\rm out}=3 \times V_{\rm out} \times M_{\rm out}/R_{\rm out} \times \tan(\alpha), \label{out-1}
\end{equation}

\noindent where $M_{\rm out}$ is the molecular gas mass of the outflow, $R_{\rm out}$ is its (projected) radial size, and  $V_{\rm out}$ is the (projected) velocity  of the outflowing gas.  We also have to assume a certain geometry given by the angle $\alpha$, which stands for the angle between the outflow and the line of sight. Equation~(\ref{out-1}) assumes a conical outflow uniformly filled by the outflowing gas (Maiolino et al.~\cite{Mai12}; Cicone et al.~\cite{Cic14}).  
 

Alternatively, in the case of a single explosive (not continuous) event, an estimate of the outflow rate can be derived from 
\begin{equation}
\dot{M}_{\rm out}=V_{\rm out} \times M_{\rm out}/R_{\rm out} \times \tan(\alpha), \label{out-3}
\end{equation}

\noindent where  the ratio  ($V_{\rm out}/R_{\rm out}$)$^{-1}$ represents the dynamical time required for the gas to reach its present location. The outflow rate from Eq.~(\ref{out-3}) is one third of the value derived from Eq.~(\ref{out-1}). 

Equation~(\ref{out-1}) was used by  Cicone et al.~(\cite{Cic14}) to study the properties of the molecular outflows in their compiled sample of galaxies. In the case of \object{NGC~1614}, the observed morphology of the outflow favors the biconical filled outflow scenario, and we also adopt here Eq.~(\ref{out-1}) in \object{IRAS~17208-0014} for consistency and to facilitate the comparison with the work of Cicone et al.~(\cite{Cic14}).

The kinetic luminosity ($L_{\rm out}$) and momentum rate ($\dot{P}_{\rm out}\equiv$d$P_{\rm out}$/d$t$) of the outflows are  derived, respectively, from Eqs.~(\ref{kin}) and (\ref{mom}), as follows

\begin{equation}
 L_{\rm out}=1/2 \times \dot{M}_{\rm out}  \times \left(V_{\rm out}/\cos(\alpha)\right)^2 \label{kin}
\end{equation}

\begin{equation}
\dot{P}_{\rm out}= \dot{M}_{\rm out} \times V_{\rm out}/\cos(\alpha) \label{mom} .
\end{equation}

 Table~\ref{t1} lists the values of $M_{\rm out}$ , $V_{\rm out}$, and $R_{\rm out}$ used to calculate $\dot{M}_{\rm out}$, $L_{\rm out}$, and $\dot{P}_{\rm out}$ in  \object{NGC~1614} and \object{IRAS~17208-0014}.
 The masses $M_{\rm mol}$, which include the mass of helium, were calculated from the CO data cubes by integrating the emission of the lines in the line wings and assuming a conservatively 
 low CO--to--H$_2$ conversion factor for the CO(1--0) line, which is typical of mergers: $\sim1/5$ of the MW value; i.e., we adopt $X_{\rm CO}=1/5\times2\times10^{20}$cm$^{-2}$(K~km~s$^{-1}$)$^{-1}$ 
 (Downes \& Solomon~\cite{Dow98}; Tacconi et al.~\cite{Tac08}; Daddi et al.~\cite{Dad10}; Genzel et al.~\cite{Gen10}; Bolatto et al.~\cite{Bol13a}). For  \object{IRAS~17208-0014}, we adopt a 
 2--1/1--0 brightness temperature ratio of 1 to derive  $M_{\rm out}$ from the CO(2--1) integrated flux.

The implied outflow rates given by Eq.~(\ref{out-1}) are $\dot{M}_{\rm out} \sim 62 \times \tan(\alpha)$~M$_{\sun}$~yr$^{-1}$ in \object{NGC~1614}, and $\sim  510\times \tan(\alpha)$~M$_{\sun}$~yr$^{-1}$ in \object{IRAS~17208-0014}. 
Assuming that the outflow in \object{NGC~1614} is perpendicular to the main disk, a geometry suggested by the observed kinematics discussed in Sect.~\ref{1614-ch}, $\alpha=i=33^{\circ}$, then $\dot{M}_{\rm out} \sim 40$~M$_{\sun}$~yr$^{-1}$. We have discarded the coplanar solution for the \object{IRAS~17208-0014} outflow ($\alpha=90^{\circ}-i=35^{\circ}$, with $i=55^{\circ}$) in Sect.~\ref{17208-mod},  but its geometry is less well constrained. The observed kinematics
nevertheless suggest that the approaching cone is not perpendicular  to the disk ($\alpha<i=55^{\circ}$).  We therefore adopt $35^{\circ}$  in the following as an indicative upper limit to $\alpha$ in \object{IRAS~17208-0014}, which implies that $\dot{M}_{\rm out} \sim 330$~M$_{\sun}$~yr$^{-1}$. 


 \begin{figure*}[tbh!]
   \centering  
  \includegraphics[width=18cm]{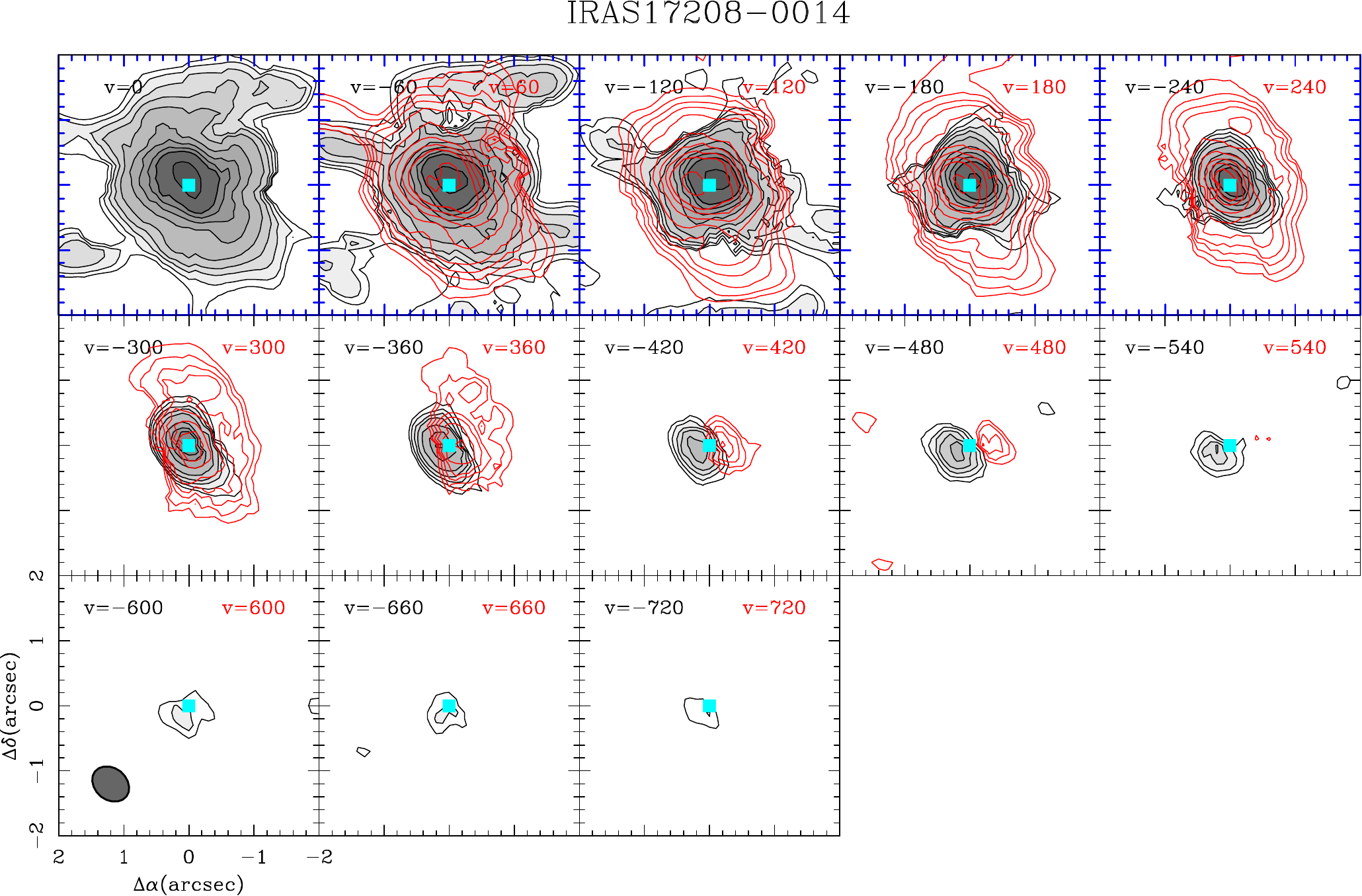}
    \caption{CO(2--1) velocity-channel maps of \object{IRAS~17208-0014}  obtained after subtracting the rotation curve model, as discussed in Sect.~\ref{17208-extreme}. 
    Contours, symbols, and scales are the same as in Fig.~\ref{IRAS17208-channels}. The dividing line between the line core and the line wing is here shifted to  $\mid$$v-v_{\rm sys}$$\mid$$~\sim300$~km~s$^{-1}$.}
    
                \label{IRAS17208-velres}
\end{figure*}

 \begin{figure*}[tbh!]
   \centering  
  \includegraphics[width=17cm]{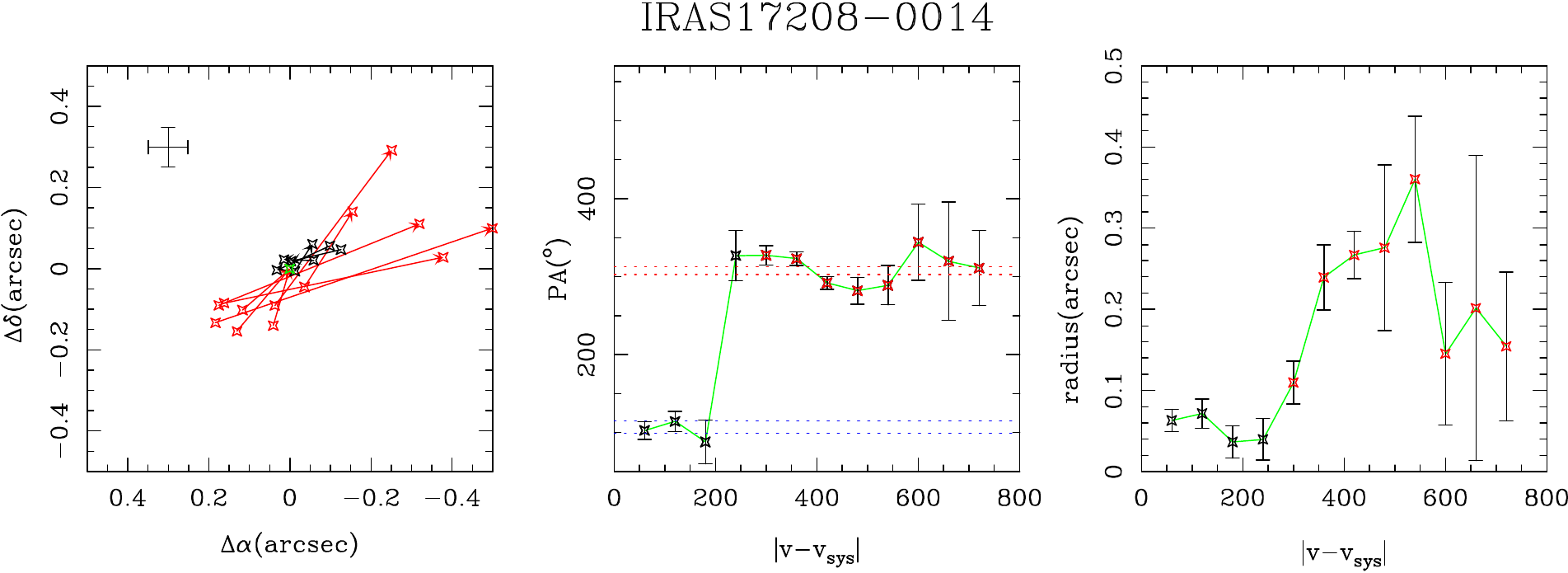}  
  
       \caption{Same as Fig.~\ref{PA-dist-17208} but for the centroids of CO(2--1) emission in \object{IRAS~17208-0014} derived from the data cube after subtraction of the rotation curve.  Red is used  for velocity channels of the line wings as redefined in Sect.~\ref{17208-extreme}. Black is used for the line core.}
              \label{PA-dist-17208-velres}
\end{figure*}
   

\subsection{The powering sources of the molecular outflows: SF or AGN -driven?} \label{out-power}

To investigate what the likely  drivers of the observed molecular outflows are (star formation and/or AGN feedback), we derived the mass, energy, and momentum loading factors. These are defined, respectively, as: 1) the ratio of $\dot{M}_{\rm out}$  to the integrated $SFR$ ($\dot{M}_{\rm out}/SFR$), 2) the ratio of  $L_{\rm out}$ to the estimated luminosity of the AGN ($L_{\rm out}/L_{\rm AGN}$) and to the total kinetic luminosity injected by supernova explosions ($L_{\rm out}/L_{\rm SNe}$), and 3) the ratio of  $P_{\rm out}$  to the total momentum rate transferred either by the AGN photons ($\dot{P}_{\rm out}/(L_{\rm AGN}/c)$) or by all the photons  ($\dot{P}_{\rm out}/(L_{\rm bol}/c)$). The kinetic luminosity due to supernovae is estimated as $L_{\rm SNe}$(erg~s$^{-1}$)~$\sim7\times10^{41}~SFR~(M_{\sun}$yr$^{-1}$)(Veilleux et al.~\cite{Vei05}). We evaluate $L_{\rm bol}\sim L_{\rm IR}$ in \object{NGC~1614} and $L_{\rm bol}\sim 1.15 \times L_{\rm IR}$ in \object{IRAS~17208-0014}, in line with the prescriptions adopted by Veilleux et al.~(\cite{Vei09}) and Cicone et al.~(\cite{Cic14}) for luminous and ultraluminous starbursts. 

We have explored a range of values for $L_{\rm AGN}$. In \object{NGC~1614} Herrero-Illana et al.~(\cite{Her14}) estimated an upper limit to the non-detected AGN in this source $L_{\rm AGN}/L_{\rm bol} < 10\%$, i.e.,  $L_{\rm AGN}<4.5\times10^{10}$~L$_{\sun}$. In the following we nevertheless adopt the more restrictive upper limit to the AGN luminosity in \object{NGC~1614} derived by Pereira-Santaella et al.~(in prep.), who combined the continuum observations at 435$\mu$m of ALMA (Xu et al.~\cite{Xu15}) and  the high-resolution  MIR observations (at 24.5$\mu$m) done with the CanariCam instrument of the GTC, to do a SED fitting of the nuclear emission in a $0\farcs5$-aperture: $L_{\rm AGN}<1.3\times10^{10}$L$_{\sun}\sim3\%L_{\rm bol}$. In \object{IRAS~17208-0014}, the values discussed in the literature for $L_{\rm AGN}$ span a wide range from $L_{\rm AGN}/L_{\rm bol} < 0.01\%$, i.e., $L_{\rm AGN}<2.4\times10^{9}$~L$_{\sun}$, as determined by Gonz\'alez-Mart\'{\i}n et al.~(\cite{Gon09}), based on X-ray observations, up to $L_{\rm AGN}/L_{\rm bol} \sim 0.3$; i.e., $L_{\rm AGN}\sim7.2\times10^{11}$L$_{\sun}$, as derived by Aalto et al.~(\cite{Aal15b}),
based on the modeling of the vibrationally excited HCN emission detected in the nucleus. Table~\ref{t1} lists the values obtained for the loading factors as defined above.

\begin{figure*}[tbh!]
   \centering  
  \includegraphics[width=6cm]{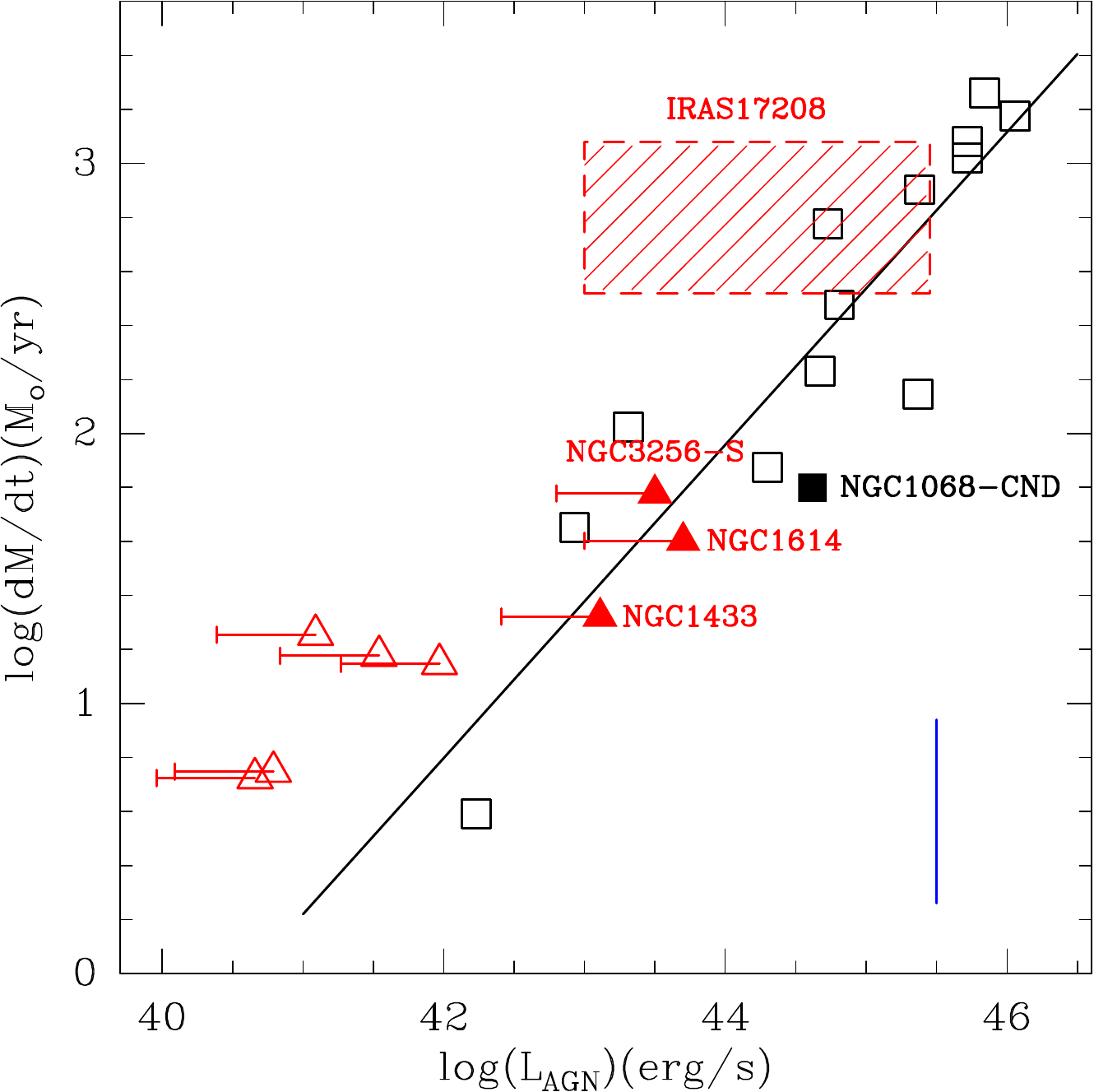}
  \includegraphics[width=6cm]{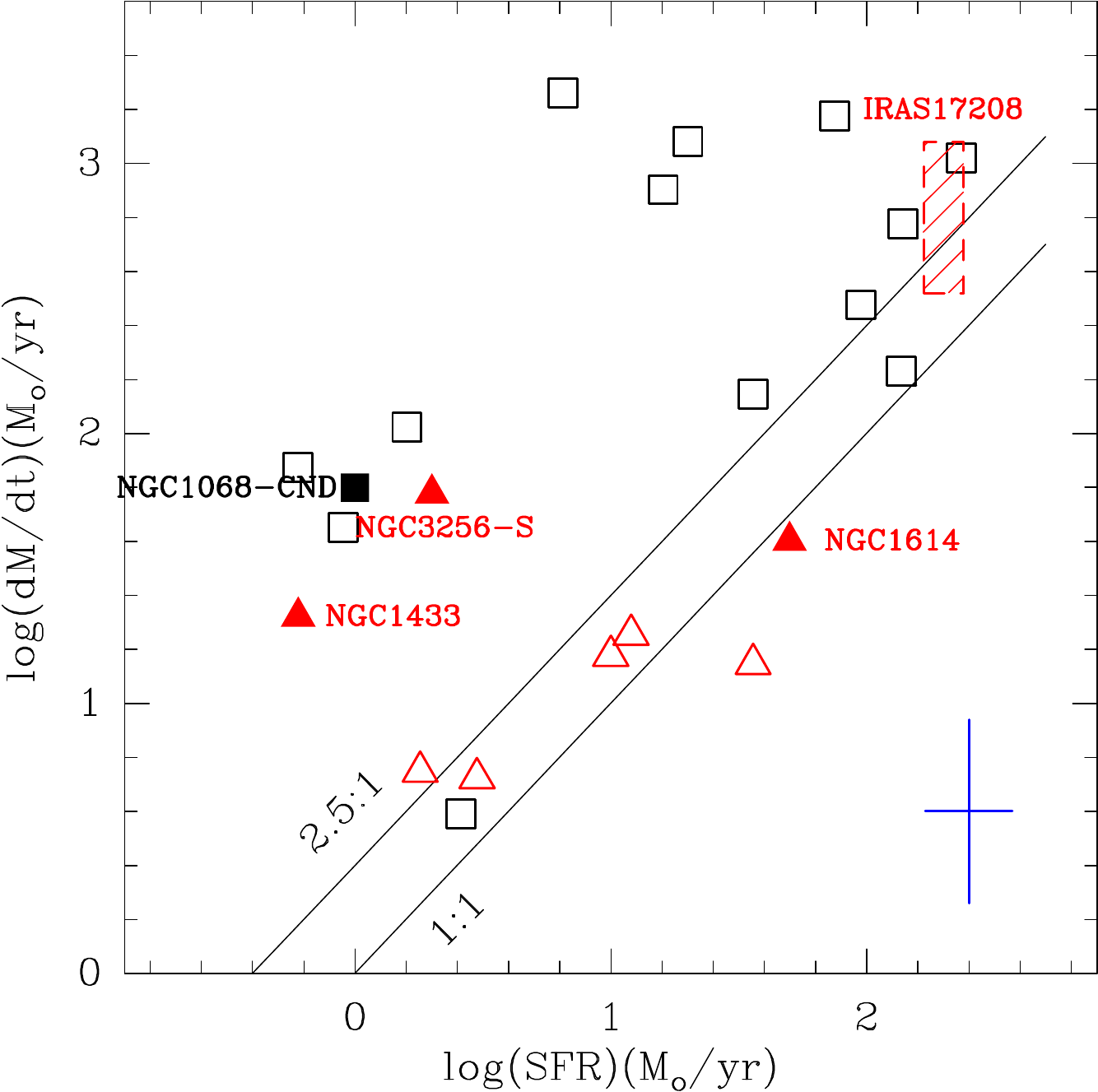}
  \includegraphics[width=6cm]{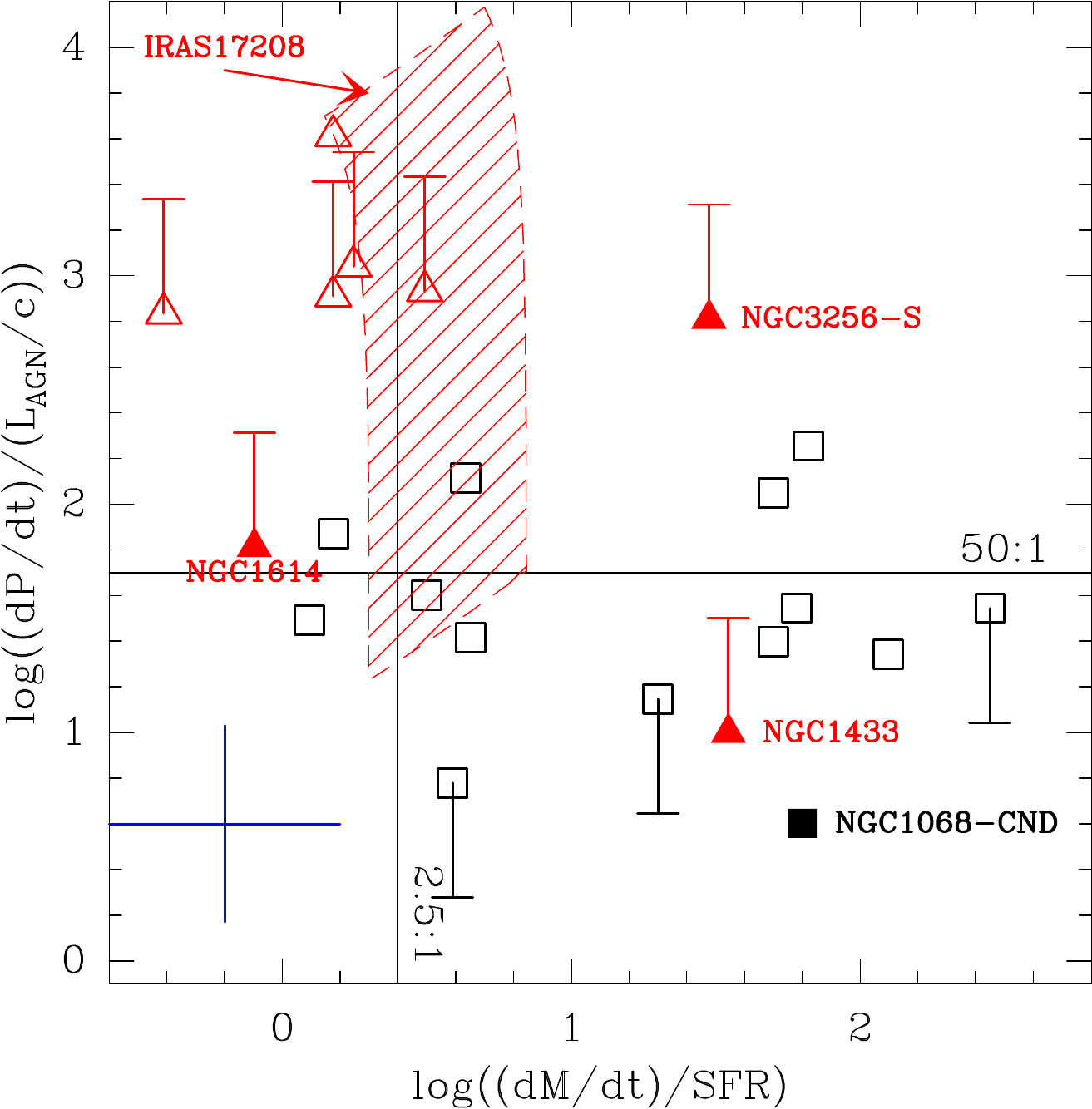}
       \caption{Scaling relations derived for the molecular outflows observed in the sample of galaxies compiled for this work: outflow mass-loss rate as a function of the AGN bolometric luminosity ({\it left panel}),  outflow mass-loss rate as a function of the SFR  ({\it middle panel}), and momentum boost factor as a function of mass-loading factor ({\it right panel}). Open symbols identify galaxies from the sample of Cicone et al.~(\cite{Cic14}). Open (red) triangles show the data from pure starburst galaxies; open (black) squares identify the data of low and high-luminosity AGNs, where the presence of an AGN has been securely detected. Filled symbols identify the data from the new additions discussed in Sect~\ref{out-comparison}: NGC~1433 (Combes et al.~\cite{Com13}),  the CND of NGC~1068~$\equiv$~NGC~1068-CND (Garc\'{\i}a-Burillo et al.~\cite{Gar14}), the southern nucleus of NGC~3256~$\equiv$~NGC~3256-S (Sakamoto et al.~\cite{Sak14}; Emonts et al.~\cite{Emo14}), as well as \object{NGC~1614} and \object{IRAS~17208-0014} (this work). The hashed polygons show the allowed range of values for \object{IRAS~17208-0014} in all the panels. The line in the {\it left panel}  shows the linear fit to the AGN sample. Lines in the {\it middle} and {\it right panels} illustrate different values of the mass and momentum loading factors. One-sided errorbars identify upper and lower limits on the values plotted in all the panels. In the bottom right of the {\it left} and {\it middle  panels} and in the bottom left of the {\it right panel} we show 1$\sigma$ errorbars that are representative of the whole sample used in this work (see discussion in Sect.~\ref{out-power} and Cicone et al.~\cite{Cic14}).}
              \label{outflow-scalings}
\end{figure*}

 We have evaluated the uncertainties associated with the quantities listed in Table~\ref{t1} (and also those listed in Table~\ref{t2} and the related Fig.~\ref{outflow-scalings}).  Outflow mass rates, momentum rates, and  kinetic luminosities have estimated uncertainties of $\pm0.34$~dex, $\pm0.36$~dex, and  $\pm0.40$~dex, respectively. These include the $\pm0.3$~dex uncertainties on the mass estimates caused by calibration and conversion factors, as well as the  errors on the sizes, the inclinations, and the velocities of the outflow, contributing $\pm0.1$~dex each. The corresponding mass, momentum, and energy loading factors have associated uncertainties of $\pm0.43$~dex, assuming that uncertainties on $SFR$ and $L_{\rm SNe}$, derived from  $L_{\rm IR}$,   are of  $\pm0.17$~dex. The errors on $L_{\rm AGN}$ (and those derived for the magnitudes that depend on them) are reflected in the upper limit and wide range of values available for \object{NGC~1614} and \object{IRAS~17208-0014}, respectively.
  Overall, these error estimates are in line with the ones commonly derived in the literature for observations of molecular outflows similar to those presented in this work  (see, e.g., Cicone et al.~\cite{Cic14}).

 An inspection of the results listed in Table~\ref{t1} leads to the following conclusions regarding the nature of the powering  sources of the outflows in \object{NGC~1614} and \object{IRAS~17208-0014}.

{\it ---\object{NGC~1614}:} 

The bulk of the total  $SFR$ comes from the nuclear ring ($\sim 50~M_{\sun}$yr$^{-1}$; Alonso-Herrero et al.~\cite{Alo01}; U et al.~\cite{U12}).  In this region, where the molecular outflow is detected, $SFR$ is comparable to $\dot{M}_{\rm out}$. Furthermore, the energy and momentum requirements of the outflow, measured by $L_{\rm out}/L_{\rm SNe}$ and  $\dot{P}_{\rm out}/(L_{\rm bol}/c)$, can be met by the star formation activity of the ring, assuming a conservatively low coupling efficiency ($\sim6\%$) for the energy injected by supernovae explosions and a very low momentum boost
($\sim 1.9$) for all the photons (which come mainly from star formation). Similarly, based on the most restrictive  limit on $L_{\rm AGN}$, the putative AGN in \object{NGC~1614} could also drive the outflow, although in this scenario the momentum boost of AGN photons should be $\geq60$. This factor is nevertheless at the higher end of the  range of values  predicted by AGN feedback models under the assumption that molecular outflows are energy-conserving: $\dot{P}_{\rm out}/(L_{\rm AGN}/c)\sim10-50$ (Faucher-Gigu\`ere \& Quataert~\cite{Fau12}). In summary, both star formation and AGN activity could cooperate to drive the molecular outflow in this source. The geometry of the line wing emission described in Sect.~\ref{1614-ch} indicates that the outflow axis is nearly perpendicular to the rotating disk, and it is spatially extended. While we cannot exclude that an AGN can also contribute to launch the outflow in a direction perpendicular to the disk, that the region in the disk where the outflow is detected extends on scales $\sim0.5-1$~kpc suggests that star formation might be its main driving agent.



{\it ---\object{IRAS~17208-0014}:} 

 As in \object{NGC~1614}, the integrated $SFR\sim 240~M_{\sun}$yr$^{-1}$, derived from a star-formation-dominated IR luminosity using a Chabrier IMF and the prescription of Kennicutt~(\cite{Ken98}), is comparable to $\dot{M}_{\rm out}$. Furthermore, the energy requirements of the outflow can only be met by assuming a significantly high coupling efficiency ($\sim40\%$) for the energy injected by supernovae. The momentum boost factor is nevertheless moderate ($\sim 5$) if all photons are considered to contribute. 
In this scenario where most of the bolometric luminosity comes from star formation ($L_{\rm AGN}<2.4\times10^{9}~$L$_{\sun}$; Gonz\'alez-Mart\'{\i}n et al.~\cite{Gon09}), the AGN power falls short of 
accounting for the luminosity and momentum rate of the molecular outflow by large factors. However, if the limit on the AGN luminosity is as high as $\sim30\%$ of $L_{\rm bol}$ (Aalto et al.~\cite{Aal15b}), i.e., a factor of three higher than estimated from MIR diagnostics by Rupke \& Veilleux~(\cite{Rup13}), this hidden AGN would be able to explain the molecular outflow with a low coupling efficiency, $L_{\rm out}/L_{\rm AGN} \sim 0.02$, and a moderate momentum boost factor, $\dot{P}_{\rm out}/(L_{\rm AGN}/c)\sim17$.  The geometry of the line wing emission discussed in Sect.~\ref{17208-ch}, which indicates that the outflow axis is not perpendicular to the large-scale rotating structure,  suggests that the outflow is launched by a non-coplanar disk that hosts either a compact starburst and/or an AGN. A merger episode could explain a {\it random} orientation of the inner circumnuclear gas disk relative to the larger scale disk. We describe in Sect.~\ref{17208-extreme} an alternative (less restrictive) definition of the velocities of the outflow in \object{IRAS~17208-0014} and the implications thereof.


\begin{table*}[bt!]
\caption{\label{t2} Re-evaluation of the molecular outflow properties in \object{IRAS~17208-0014}}
\centering
\begin{tabular}{lccccccccccc}
\noalign{\smallskip} 
\hline
\hline
\noalign{\smallskip} 
Source & $R_{\rm out}$  & $V_{\rm out}$ & $M_{\rm out}$ & $\dot{M}_{\rm out}$ & $ \frac{\dot{M}_{\rm out}}{SFR}$      & $L_{\rm out}$ & $\frac{L_{\rm out}}{L_{\rm SNe}}$ &  $\frac{L_{\rm out}}{L_{\rm AGN}}$ & $\dot{P}_{\rm out}$ &     $\frac{\dot{P}_{\rm out}}{L_{\rm bol}/c}$ & $\frac{\dot{P}_{\rm out}}{L_{\rm AGN}/c}$        \\
\noalign{\smallskip}
              &  (pc) & (km~s$^{-1}$)   &       ($M_{\sun}$)        &   ($M_{\sun}$~yr$^{-1}$)  &       --      &       ($L_{\sun}$)    & --      &    -- &       (g~cm~s$^{-2}$) & -- & -- \\
\noalign{\smallskip}                  
\hline
\noalign{\smallskip} 
 \object{IRAS~17208}    & 160  &        500     & $2\times10^{8}$          &     1200    &       $5~$\tablefootmark{a}   &       $4\times10^{10}$ & $1~$\tablefootmark{a} &       $>15~$\tablefootmark{b} & $4.8\times10^{36}$   &  15    & $>1.5\times10^{4}~$\tablefootmark{b}       \\
        &  -- & --      &  --   &  --   & $7~$\tablefootmark{c}  & -- & $1.4~$\tablefootmark{c}         &       $0.05~$\tablefootmark{d}        &  --  &   --   & $51~$\tablefootmark{d}         \\
 \noalign{\smallskip}    
\hline
\hline 
\end{tabular} 
\tablefoot{New properties derived for the line wings as defined in Sect.~\ref{17208-extreme} assuming $\alpha=35^{\circ}$. 
Outflow mass rates, momentum rates, and  kinetic luminosities have uncertainties of $\pm0.34$~dex, $\pm0.36$~dex, and  $\pm0.40$~dex, respectively. The mass, momentum, and energy loading factors have associated uncertainties of $\pm0.43$~dex. See Sect.~\ref{out-power} for details. 
\\
\tablefoottext{a}{Derived assuming $SFR=240~M_{\sun}$~yr$^{-1}$.}  \\
\tablefoottext{b}{Derived assuming $L_{\rm AGN}^{\rm IRAS~17208}<2.4\times10^{9}~$L$_{\sun}$     (Gonz\'alez-Mart\'{\i}n et al.~\cite{Gon09}).}\\
\tablefoottext{c}{Derived assuming $SFR=0.7\times240=168~M_{\sun}$~yr$^{-1}$.}  \\
\tablefoottext{d}{Derived assuming $L_{\rm AGN}^{\rm IRAS~17208}\sim7.2\times10^{11}~$L$_{\sun}$    (Aalto et al.~\cite{Aal15b}).}}\\
\end{table*}


\subsection{An extreme AGN-driven outflow in \object{IRAS~17208-0014}?} \label{17208-extreme}

The molecular outflow axis in \object{IRAS~17208-0014} is significantly tilted relative to the kinematic major axis of the rotating disk, as discussed in Sect.~\ref{17208-kin}. To evaluate the intrinsic velocity boost in the outflow relative to the rotating frame of the disk in \object{IRAS~17208-0014}, we obtained a new version of the data cube by subtracting the rotation curve model derived in Sect.~\ref{17208-kin}\footnote{We note that applying the same correction to  the data cube of \object{NGC~1614} (results are not shown here) had no significant effect on the velocity pattern of the outflow because the latter is oriented close to the minor axis of the rotating disk, as discussed in Sect.~\ref{1614-kin}.} at each pixel.

Figure~\ref{IRAS17208-velres} shows the same velocity-channel maps as are displayed in  Fig.~\ref{IRAS17208-channels} but after subtracting the rotation curve.  This figure highlights that the decoupling of 
the outflow component relative to the rotating disk, which is revealed by the apparent {\em \emph{reversal}} of the respective kinematic major axes, is identified by visual inspection at  
$\mid v-v_{\rm sys}\mid$~$\geq300$~km~s$^{-1}$. This result significantly expands the velocity range of the line wings, here defined relative to the rotating frame of the disk, compared to the description of Sect.~\ref{17208-ch}.  
In particular, the redshifted counterpart of the outflow, which remained {\em \emph{unnoticed}} in  Fig.~\ref{IRAS17208-channels}, is now identified  in the range $v-v_{\rm sys}$~=~[+300, +540]~km~s$^{-1}$ (see 
Fig.~\ref{IRAS17208-velres}).

To quantify the effect of the decoupling, we show the centroids  of CO emission in \object{IRAS17208-0014} in  Fig.\ref{PA-dist-17208-velres},  as well as  the $PA$ and $r$ 
profiles obtained for the velocity channels shown in Fig.~\ref{IRAS17208-velres}.  As illustrated by the middle panel of Fig.~\ref{PA-dist-17208-velres}, the value of $PA$ for  $\mid v-v_{\rm sys}\mid$~$\leq200$~km~s$^{-1}$  
is $PA_{\rm core}=107\pm8^{\circ}$, i.e., close to the $PA$ of the rotating disk derived in Sect.~\ref{17208-kin}. This indicates that the rotation pattern is still present in this velocity range of the {\it \emph{kinematically rescaled}} data, reflecting that 
the model underestimates the intrinsic  rotation curve, as already mentioned in Sect.~\ref{17208-outflow}. In stark contrast, the $PA$ values for the centroids at 
$\mid v-v_{\rm sys}\mid$~$\geq300$~km~s$^{-1}$
oscillate around a weighted mean of $PA_{\rm wing}=308^{\circ}\pm5^{\circ}$, revealing a true reversal of the kinematic major axis of the outflow relative to the disk.  We conclude that the evidence of
a kinematic decoupling, reflected in the measured difference $PA_{\rm wing}-PA_{\rm core}\sim201\pm9^{\circ}$ is a significant $\geq20\sigma$ result. 
Furthermore, the line wing channels show slightly lower $r$ values ($\sim0\farcs2$~($160$~pc)) compared to the   line core defined in Sect.~\ref{17208-ch} (up to $\sim0\farcs3$~(240~pc); see the right-hand panel of Fig.~\ref{PA-dist-17208}).

Table~\ref{t2} lists the new values obtained for the loading factors of the outflow in \object{IRAS~17208-0014} derived from the redefined velocity range. The enhanced mass loading factor ($\dot{M}_{\rm out}/SFR\sim 5-7$) cannot be easily explained by star formation activity alone.  From the observational point of view, mass loading factors exceeding unity are rarely seen in actively star-forming mergers, in particular in  the range $SFR>100~M_{\sun}$~yr$^{-1}$ applicable to  \object{IRAS~17208-0014} (Rupke et al.~\cite{Rup05b}; Sturm et al.~\cite{Stu11}). Moreover, Table~\ref{t2} shows that the required efficiency for converting the energy injected by supernovae into mechanical energy of the outflow would be unrealistically high ($\geq 100\%$).  Characteristic values defined observationally for the energy conversion efficiency oscillate between 10$\%$ and, at most,  50$\%$ (Veilleux et al.~\cite{Vei05}; Veilleux~\cite{Vei08}).  By contrast, a putative hidden AGN of $L_{\rm AGN}\sim7.2\times10^{11}~$L$_{\sun}$ would be a viable driving agent of the molecular outflow in this source. This scenario would require a moderate coupling efficiency, $L_{\rm out}/L_{\rm AGN} \sim 0.05,$ and an admittedly high momentum boost factor, $\dot{P}_{\rm out}/(L_{\rm AGN}/c)\sim51$. These values are nevertheless in line with those commonly derived in other AGN-driven outflows (Cicone et al.~\cite{Cic14}).

Independent evidence of a possible {\it \emph{hidden}} AGN in  \object{IRAS~17208-0014} can be found in the new ALMA observations of the ($v2=1$) 4--3 HCN 
vibrational line emission (Aalto et al.~\cite{Aal15b}). \object{IRAS~17208-0014} is part of a family of (U)LIRGs where the emission in vibrationally excited lines 
seems to emerge from  buried, compact ($r<17-70$~pc) nuclei that have very high implied mid-infrared surface brightness $>5\times10^{13}L_{\sun}$~kpc$^{-2}$. These nuclei are likely powered by accreting supermassive black holes (SMBHs) and/or hot ($>200$~K) extreme starbursts. The reported extreme surface brightness of the hidden core in  \object{IRAS~17208-0014} would be able to  excite the HCN vibrational line by intense 14~$\mu$m emission.
The new observations of \object{IRAS~17208-0014} presented by Aalto et al.~(\cite{Aal15b}) also show evidence of an extreme outflow  in the dense molecular gas traced by the ground vibrational state ($v=0$) HCN(4--3) line emission imaged by ALMA.

\section{The molecular outflows of  \object{NGC~1614} and \object{IRAS~17208-0014} in context}\label{out-comparison}

We compiled a sample of 23 nearby galaxies where molecular outflows have been observed and detected to date, taking the sample of 18 objects studied by 
Cicone et al.~(\cite{Cic14}) as starting point, complemented here by new data recently obtained in five targets. These additions include two Seyferts (NGC~1433: Combes et al.~\cite{Com13} and the CND of NGC~1068~$\equiv$~NGC~1068-CND: Garc\'{\i}a-Burillo et al.~\cite{Gar14}) and three mergers (the southern nucleus of NGC~3256~$\equiv$~NGC~3256-S: Sakamoto et al.~\cite{Sak14} and Emonts et al.~\cite{Emo14}, as well as \object{NGC~1614} and \object{IRAS~17208-0014}: this work). We have also included the new outflow parameters derived by Alatalo et al.~(\cite{Ala15}) in NGC~1266.  Figure~\ref{outflow-scalings} illustrates the different scaling relations (on logarithmic scale) obtained in an attempt to identify the driving agents of the molecular outflows inside the parameter space probed by the sample.

The new data of NGC~1433, NGC~1068-CND, NGC~1614, and NGC~3256-S follow the correlation between the outflow rate and the AGN luminosity  found by Cicone et al.~(\cite{Cic14}) in their sample, which is an indication that the molecular outflow in these sources can be AGN-driven (left-hand panel of Fig.~\ref{outflow-scalings}). The location of \object{IRAS~17208-0014} in the $SFR-L_{\rm AGN}$ diagram depends critically on the wide range of values for the AGN power discussed in Sect.~\ref{out-power}. This source follows the correlation if the highest value of  $L_{\rm AGN}$ is adopted but it becomes an outlier, like the pure starbursts of Cicone et al.~(\cite{Cic14}), if the low end of $L_{\rm AGN}$ is assumed. 

The middle panel of Fig.~\ref{outflow-scalings} shows the outflow rate as a function of the SFR. In this diagram,  NGC~1433, NGC~1068-CND, and NGC~3256-S share the location of other AGN-driven outflows as they all depart from the 1:1 line followed by pure starbursts. In contrast, NGC~1614 follows the correlation of pure starbursts, an indication that star formation can also cooperate to drive the outflow in this source. The location of \object{IRAS~17208-0014} in this diagram depends on the value adopted for $\dot{M}_{\rm out}$: the more extreme loading factors discussed in Sect.~\ref{17208-extreme} put it beyond the 2--3:1 line predicted by numerical simulations that include radiative and mechanical feedback in star-forming galaxies (Hopkins et al.~\cite{Hop12}; Lagos et al.~\cite{Lag13}).

The right-hand panel of Fig.~\ref{outflow-scalings} shows the momentum boost factor as a function of the mass-loading factor for the compiled sample. Galaxy outflows in the upper left-hand (lower right-hand) quadrant with momentum-boost factors $>50$ ($<50$) and mass-loading factors $<2.5$ ($>2.5$) qualify as star formation (AGN) driven systems. Galaxies lying in the lower left-hand quadrant are mixed systems to the extent that both star formation and AGN can cooperate to drive the outflow in these sources. \object{NGC~1614} is close to this {\it \emph{undefined}} region.  Galaxies like NGC~3256-S, which lies in the upper right-hand quadrant, likely require other mechanisms like radio jets to drive the outflow. The location of \object{IRAS~17208-0014} in this diagnostic diagram reflects a range of possible classifications. However, if we assume the properties derived in Sect.~\ref{17208-extreme}, a hidden AGN is the most likely powering source of the outflow in this source.

\section{Summary and conclusions}\label{Summary}

  We have used the PdBI to image the CO emission in two-starbursts dominated (U)LIRGs:  \object{NGC~1614} and \object{IRAS~17208-0014} with high spatial resolution (0$\farcs$5--1$\farcs$2). The two targets, classified as mergers in an advanced stage of interaction, are vigorous star-forming systems ($SFR\geq50-240~M_{\sun}$~yr$^{-1}$), and they present evidence of an outflow in several ISM phases, including the ionized and the neutral gas.  The high resolution of the PdBI observations used in this work are instrumental in constraining the geometry, as well as the mass, momentum, and energy budgets of the molecular outflows discovered in \object{NGC~1614} and \object{IRAS~17208-0014}.

  We summarize the main results of our study as follows:
  
  \begin{itemize}
  
  \item

  The CO(1--0) emission in \object{NGC~1614} has been mapped with a spatial resolution of $1\farcs5\times0\farcs8$. The bulk of the CO emission comes from a spatially resolved  (3.4~kpc~$\times$~2.5~kpc), lopsided molecular disk. The disk shows an arc-like feature that runs from the northeastern to the southeastern side.  A sizable fraction of the CO emission originates in the 600~pc--diameter star-forming ring of supergiant H{\small II} regions, which is associated with a young starburst ($SFR\sim50~M_{\sun}$~yr$^{-1}$). Fainter CO emission is also detected and extends farther out to the east up to $r\sim1.9$~kpc.

   \item

 The CO emission in \object{NGC~1614} extends over a symmetric $\sim900$~km~s$^{-1}$-wide velocity span. The bulk of the emission comes from a more restricted  velocity range: $\mid v-v_{\rm sys} \mid~ \leq 210$~km~s$^{-1}$ (the line core). We also detect emission that extends up to $\mid v-v_{\rm sys} \mid~\sim 450$~km~s$^{-1}$ (the line wings). Emission from the line core shows the signature of rotating disk with a kinematic major axis oriented roughly north-south, as confirmed by  a Fourier decomposition of the mean velocity field.  In contrast, the kinematic major axis for the line wings is tilted progressively toward the east-west axis, an indication of strong kinematic decoupling of the high-velocity gas.  The kinematic pattern of the line wings can be explained by a non-coplanar molecular outflow. 
 
 \item
 
 The molecular outflow in  \object{NGC~1614} has a mass of $3.2\times10^{7}M_{\sun}$ and consists mainly of two knots of emission that are co-spatial with the nuclear star-forming ring. The eastern (western) knot is blue(red)-shifted by $\sim$360~km~s$^{-1}$ on average. The blueshifted velocity ranges of the outflow are comparable in CO and H$\alpha$.  The mass, energy, and momentum budget requirements of the molecular outflow in \object{NGC~1614} can be met by the star formation activity taking place in the ring. The putative AGN in \object{NGC~1614} could also drive the outflow. The geometry of the line wing emission indicates that the outflow axis is nearly perpendicular to the rotating disk, and it is spatially extended on scales $\sim0.5-1$~kpc. This  suggests that star formation is the most likely driving agent of the outflow, although a contribution from a putative AGN cannot be excluded.

  \item

  The CO(2--1) emission in \object{IRAS~17208-0014} has been mapped with a spatial resolution $0\farcs6\times0\farcs5$.  The CO emission comes from a spatially resolved  (2.7~kpc~$\times$~1.8~kpc) molecular disk. The morphology of the CO disk is similar to the northeast-southwest elongated structure seen in the dust extinction, as derived from the 2.2$\mu$m/1.1$\mu$m HST color images.   Three fainter protrusions and a detached clump extend the disk emission farther out up to $r\sim1.8$~kpc. The molecular disk feeds a vigorous star formation episode  ($SFR\sim240~M_{\sun}$~yr$^{-1}$). 
 
 \item
  The CO emission in \object{IRAS~17208-0014}  covers a wide and asymmetric velocity range: $v-v_{\rm sys}$~=~[+450,--750]~km~s$^{-1}$. However, up to $\simeq 99\%$ of the total emission comes from a narrower velocity range: $v-v_{\rm sys} \sim [+450,-450]$~km~s$^{-1}$ (the line core) that shows the characteristic feature of spatially resolved rotating disk with 
   $PA=113\pm3^{\circ}$, as determined by  a Fourier decomposition of the mean velocity field. Rotation is perturbed by strong non-circular motions.  Gas emission from the blueshifted line wing  is kinematically decoupled from the disk: its apparent  kinematic major axis is virtually reversed. This pattern can be explained by a non-coplanar molecular outflow. This scenario is reinforced after subtracting the rotation curve, which reveals that the kinematic decoupling of the outflow starts at $\mid v-v_{\rm sys}\mid$~$\geq300$~km~s$^{-1}$.

   \item
The molecular outflow in  \object{IRAS~17208-0014} has a mass of $2\times10^{8}M_{\sun}$ and consists of two knots of emission located at $r\sim160$~pc. The southeast (northwest) knot is blue(red)-shifted by $\sim$500~km~s$^{-1}$ on average. The blueshifted velocity ranges of the outflow are similar in atomic, ionized, and molecular gas.  In stark contrast to \object{NGC~1614},  the mass, energy, and momentum budget requirements of the molecular outflow in \object{IRAS~17208-0014} cannot be met by its star formation activity. Instead, a putative hidden AGN of $L_{\rm AGN}\sim7.2\times10^{11}~$L$_{\sun}$ would be a viable driving agent of the molecular outflow in this source. The geometry of the molecular outflow, with an axis that is not perpendicular to the large-scale rotating structure,  suggests that the outflow is launched by a non-coplanar disk.

 \end{itemize}

  The molecular outflow  in \object{NGC~1614} is likely triggered by its vigorous star formation 
activity, which is mainly concentrated in the 600~pc--diameter nuclear ring of the galaxy. This explains the extended nature of the outflow, but also its orientation, which is roughly orthogonal relative to the large-scale rotating disk.
In the case of  \object{IRAS~17208-0014,} the molecular outflow geometry and energetics can be explained best as driven by an AGN.  It is tempting to identify which of the two overlapping 
nuclear stellar disks discovered in  \object{IRAS~17208-0014}, which lie 0\farcs2 (200~pc) apart and which were recently discovered by Medling et al.~(\cite{Med14}), is responsible for launching the outflow in this source. Although our spatial resolution (0\farcs5) is insufficient to precisely locate the origin of the outflow spatially, the observed kinematics can be used to make a tentative identification.     
Medling et al.~(\cite{Med14}) characterize the two rotating stellar disks kinematically: the major axis is oriented at $PA\sim40^{\circ}$ for the W nucleus and at  $PA\sim128^{\circ}$ for 
the E nucleus. While the $PA$ of the large-scale rotating molecular disk ($PA\sim113^{\circ}$) is closer to the orientation of the E disk, the derived molecular outflow axis ($PA\sim308^{\circ}$) is, within a $2^{\circ}$ tolerance, perpendicular to the W disk. This suggests that the launching mechanism of the molecular outflow in \object{IRAS~17208-0014} lies in a buried AGN associated with the W nucleus.
To be obtained with the ALMA interferometer, higher resolution observations are required to unambiguously identify the launching mechanism of the molecular outflow in \object{IRAS~17208-0014}.

\begin{acknowledgements}

We acknowledge the IRAM staff from the Plateau de Bure and from Grenoble for carrying out the observations and help provided during the data reduction. 
We used observations made
with the NASA/ESA Hubble Space Telescope, and obtained from the Hubble
Legacy Archive.
We thank Dr.~Anne Medling for sharing information on the absolute coordinates of the stellar nuclear disks of \object{IRAS~17208-0014}.
 SGB thanks the Paris Observatory for its economic support and hospitality during his stay in August 2014. SGB acknowledges support from Spanish grants AYA2010-15169 and from the Junta de Andalucia through TIC-114 and the Excellence Project P08-TIC-03531. SGB and AL acknowledge support from MICIN within program CONSOLIDER INGENIO 2010, under grant `Molecular Astrophysics: The Herschel and ALMA Era--ASTROMOL' (ref CSD2009-00038). 
 SGB, AU, LC, and SA acknowledge support from Spanish grant AYA2012-32295. 
 FC acknowledges the European Research Council for the Advanced Grant Program Num.~267399-Momentum. 
 AAH acknowledges support from the Spanish Plan Nacional grant  AYA2012-31447 (partly funded by the FEDER program).

\end{acknowledgements}



\end{document}